\documentclass[letterpaper,titlepage,11pt]{article}
\pdfoutput=1
\usepackage[colorlinks=true,urlcolor=blue,citecolor=magenta]{hyperref}
\usepackage{amssymb,amsmath,amsfonts}
\usepackage{color}
\usepackage{graphicx}
\usepackage{array}
\usepackage{epsfig}
\usepackage{epstopdf}
\usepackage{caption}
\usepackage{subcaption}
\usepackage{epsfig}
\usepackage{epstopdf}
\usepackage{eso-pic}

\newcommand\BackgroundPic{%
\put(0,0){%
\parbox[b][\paperheight]{\paperwidth}{%
\vfill
\centering
\includegraphics[width=\paperwidth,height=\paperheight,%
keepaspectratio]{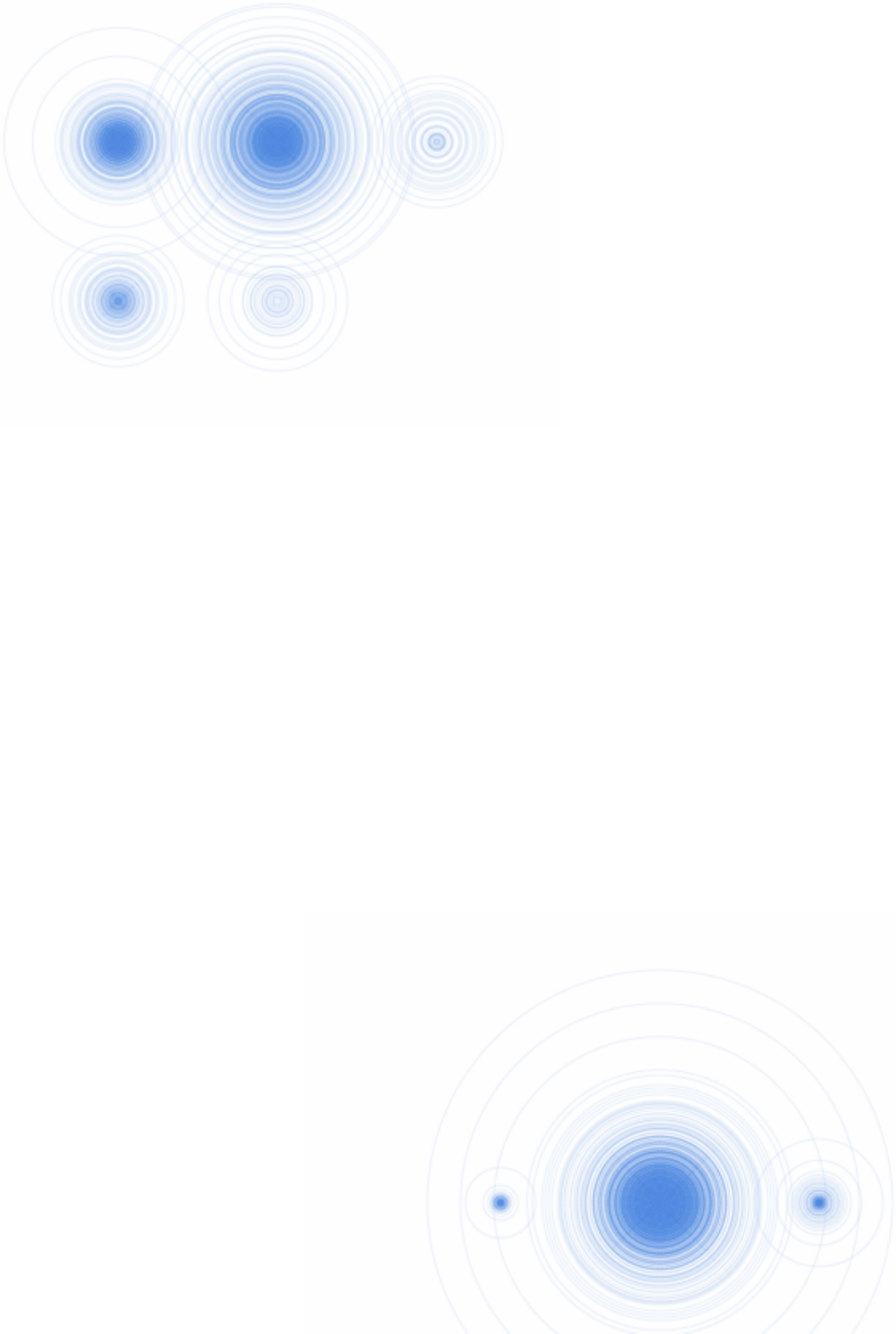}%
\vfill
}}}

\def\clock{{\count0=\time
           \divide\count0 60
           \ifnum\count0<10 0\fi\the\count0
           \multiply\count0 -60 \advance\count0 \time
           :\ifnum\count0<10 0\fi \the\count0
         }}
\newcommand{\timestamp}{{\small\vbox{\hbox{\tt\jobname.tex}
\hbox{\the\day/\the\month/\the\year, \clock}}}}

\newcommand{\beq}{\begin{equation}}
\newcommand{\eeq}{\end{equation}}

\let\oldsqrt\sqrt
\def\sqrt{\mathpalette\DHLhksqrt}
\def\DHLhksqrt#1#2{%
\setbox0=\hbox{$#1\oldsqrt{#2\,}$}\dimen0=\ht0
\advance\dimen0-0.2\ht0
\setbox2=\hbox{\vrule height\ht0 depth -\dimen0}%
{\box0\lower0.4pt\box2}}

\numberwithin{equation}{section}

\begin{document}
\AddToShipoutPicture*{\BackgroundPic}

\hypersetup{pageanchor=false}
\begin{titlepage}
 \vskip 1.8 cm

\centerline{\Huge \bf (Non)-Dissipative Hydrodynamics}
\vskip 0.5cm
\centerline{\Huge \bf on Embedded Surfaces}
\vskip 2cm

\centerline{\large {\bf Jay Armas }}

\let\thefootnote\relax\footnote{\url{http://www.jacomearmas.com} }

\vskip 1.0cm

\begin{center}
\sl  Albert Einstein Center for Fundamental Physics \\
\sl Institute for Theoretical Physics, University of Bern \\
\sl  Sidlerstrasse 5, 3012-Bern, Switzerland
\end{center}
\vskip 0.4cm

\centerline{\small\tt jay@itp.unibe.ch}

\vskip 1.3cm \centerline{\bf Abstract} \vskip 0.2cm \noindent

We construct the theory of dissipative hydrodynamics of uncharged fluids living on embedded space-time surfaces to first order in a derivative expansion in the case of codimension-1 surfaces (including fluid membranes) and the theory of  non-dissipative hydrodynamics to second order in a derivative expansion in the case of codimension higher than one under the assumption of no angular momenta in transverse directions to the surface. This construction includes the elastic degrees of freedom, and hence the corresponding transport coefficients, that take into account transverse fluctuations of the geometry where the fluid lives. Requiring the second law of thermodynamics to be satisfied leads us to conclude that in the case of codimension-1 surfaces the stress-energy tensor is characterized by 2 hydrodynamic and 1 elastic independent transport coefficient to first order in the expansion while for codimension higher than one, and for non-dissipative flows, the stress-energy tensor is characterized by 7 hydrodynamic and 3 elastic independent transport coefficients to second order in the expansion. Furthermore, the constraints imposed between the stress-energy tensor, the bending moment  and the entropy current of the fluid by these extra non-dissipative contributions are fully captured by equilibrium partition functions. This analysis constrains the Young modulus which can be measured from gravity by elastically perturbing black branes.

\end{titlepage}


\tableofcontents


\section{Introduction}
Recently, in the first study of the transport properties of stationary fluids living on submanifolds embbeded in a background space-time (fluid branes), it was shown,  using equilibrium partition function techniques, that such fluids are characterized by three sets of transport coefficients to second order in a derivative expansion that can be split into hydrodynamic, elastic and spin transport coefficients \cite{Armas:2013hsa}. Hydrodynamic transport coefficients are related to derivatives of the fluid variables and Riemann curvature terms of the embedded submanifold, while elastic transport coefficients are related to the extrinsic curvature of the submanifold and spin response coefficients to the angular momenta of the fluid in transverse directions to the submanifold.\footnote{A spinning particle moves along a worldline and is endowed with a spin-two tensor characterising its rotation along transverse planes to the worldline. Here, the spin coefficients associated to fluid branes describe the rotation of the brane in transverse planes to its worldvolume \cite{Armas:2013hsa}.} It was shown that such fluids were characterized by a family of 3 hydrodynamic, 4 elastic and 1 mixed fluid-elastic transport coefficient\footnote{Already to second order in the case of codimension-1 surfaces, though in general for any codimension to third or higher order, transport coefficients can exhibit mixed hydrodynamic, elastic and spin behaviour \cite{Armas:2013hsa}.} in the case of codimension-1 surfaces and by a family 3 hydrodynamics, 3 elastic and 1 spin transport coefficient in the case of codimension higher than one, ignoring certain dimension specific contributions \cite{Armas:2013hsa}. The corresponding entropy current analysis due to these corrections is carried out in \cite{Armas:2014rva}.

It was also shown recently that in the case of space-filling uncharged fluids (which are not confined to a submanifold), the equilibrium partition function, which only applies to fluids in stationary motion, captures 3 hydrodynamic transport coefficients \cite{Banerjee:2012iz, Jensen:2012jh}. Furthermore, by relaxing the assumption of stationarity, and appealing to symmetry arguments and the second law of thermodynamics, dissipative uncharged fluid configurations are characterized by a set of 12 hydrodynamic independent transport coefficients \cite{Bhattacharyya:2012nq} to second order in the derivative expansion.\footnote{The analysis of \cite{Bhattacharyya:2012nq} has shown that the stress-energy tensor of these fluids is characterised by a total of 2 independent transport coefficients at first order and that at second order 10 more independent transport coefficients appear.} Hence, relaxing stationarity allows for the appearance of 9 other transport coefficients. It is therefore interesting to ask \textbf{(i)} whether new dissipative elastic and spin transport coefficients appear in the case of fluids living on submanifolds if one considers non-stationary configurations and, if not, then \textbf{(ii)} are the constraints imposed by the second law of thermodynamics fully captured by the equilibrium partition function?

The motivation for answering these questions is many-fold. First of all, fluids confined to a submanifold are relevant systems for theoretical biology and soft condensed matter physics as they describe the effective dynamics of fluid membranes \cite{Helfrich1973, Canham197061, doi:10.1080/00018739700101488}. Therefore, the construction of this theory of dissipative fluid dynamics in a derivative expansion is interesting in its own right both in the relativistic and the non-relativistic cases. Secondly, there has been a large body of work in the past few years on gravitational systems dual to fluid dynamics. In particular, long wavelength fluctuations along worldvolume/boundary directions of black branes are effectively described by the dynamics of viscous fluid flows \cite{Bhattacharyya:2008jc} while perturbations along transverse directions are described by the dynamics of thin elastic branes \cite{Emparan:2007wm, Emparan:2009cs,Emparan:2009at, Armas:2012jg, Armas:2011uf, Camps:2012hw}. Worldvolume perturbations, via the gauge/gravity duality, have allowed us to gain insights into quantum field theories and furthermore, to constrain the possible structures characterizing those theories. Therefore it is interesting to try to understand whether transverse perturbations of black branes can also lead to valuable insight. Thirdly, it has been shown in different settings that the fluid configurations dual to black brane geometries need not live on the boundary of the space-time but can live in an intermediate region between the horizon and the boundary \cite{Brattan:2011my, Emparan:2013ila}. Speculating that the dynamics of such black branes may be described by more general holographic dualities in terms of a dual quantum field theory, then a generic analysis of confined fluids would constrain those theories. The goal is then to search for a complete classification of the quantities characterizing confined fluids and hence the classification of the structures, such as the stress-energy tensor and the bending moment, that can be obtained from gravity by a generic perturbation of black branes dual to uncharged fluids.

The work presented here will not fully answer the questions put forth in the beginning of this section due to several limitations that we briefly comment here and further explain during the course of this work. For codimension-1 surfaces we only construct the theory to first order in the derivative expansion. The reason for this is that in order to push one order further it would be necessary to derive the equations of motion for curved branes to pole-quadrupole order, an endeavour that is yet to be accomplished.\footnote{See Ref.~\cite{Armas:2013hsa} for a specific case of pole-quadrupole equations of motion derived from an equilibrium partition function with a mixed fluid-elastic transport coefficient.} In the case of codimension higher than one we restrict ourselves, due to the same reason, to the non-dissipative sector and construct the theory to second order in the derivative expansion but we ignore spin transport coefficients which are generically proportional to the extrinsic twist potential of the embedded submanifold. The inclusion of intrinsic spin along transverse directions to the surface in confined fluids requires a modification of the first law of thermodynamics as the intrinsic spin may be seen as a conserved $U(1)$ charge \cite{Armas:2014rva}. Before attempting such classification, one should first go through the exercise of constructing the theory of dissipative charged fluids. Therefore, we do not consider spinning fluids in the sense explained above. Given these assumptions, and some more technical ones that will be explained in Sec.~\ref{classification}, we show that the most general stress-energy tensor to first order in a derivative expansion for codimension-1 surfaces is characterized by 2 hydrodynamic and 1 elastic independent transport coefficient. In the case of codimension higher than one in the non-dissipative sector we show that the stress-energy tensor is characterized by 7 hydrodynamics and 3 elastic independent transport coefficients. Furthermore, the constraints obtained between the entropy current, the bending moment and the stress-energy tensor involving these extra transport coefficients are fully captured by equilibrium partition functions. The extra transport coefficients are thus non-dissipative, as expected from classical elasticity theory.\footnote{This is also observed in theories of viscoelastic fluids \cite{Fukuma:2011pr}.}

This work is organized as follows. In Sec.~\ref{classification} we begin by defining some properties and geometric structures associated with embedded space-time surfaces. The generic form of the equations of motion is given and the structures appearing in the equations of motion and entropy current are classified as well as the terms appearing in the divergence of the entropy current. Our assumptions in the construction of these theories are clearly stated. In Sec.~\ref{divergence} we first calculate the divergence of the entropy current and organize the several terms appearing in such operation according to the independent fluid-elastic data. Afterwards, we impose positivity of the divergence of the entropy current and solve for the constraints between the several parameters entering the entropy current, bending moment and stress-energy tensor. In Sec.~\ref{functions} we compare our results with those obtained from equilibrium partition functions. Finally, in Sec.~\ref{discussion} we summarize our work and comment on open questions and future research directions.


\section{Classification of fluid-elastic data} \label{classification}
In this section we review the necessary tools for dealing with the geometry of embedded surfaces and the tensor structures that characterize it. We then present the equations of motion that any material living on a surface must satisfy in the probe approximation when the surface is taken to have a finite thickness. These equations of motion are determined in terms of a set of tensors structures which, in order to construct a generic theory of dissipative hydrodynamics, need to be classified in terms of independent components. These components consist of all possible contributions which are allowed by symmetry and are on-shell independent. This classification is given at the end of this section and it will be the starting point for imposing the second law of thermodynamics and constraining the allowed contributions.


\subsection{Geometry of embedded surfaces}
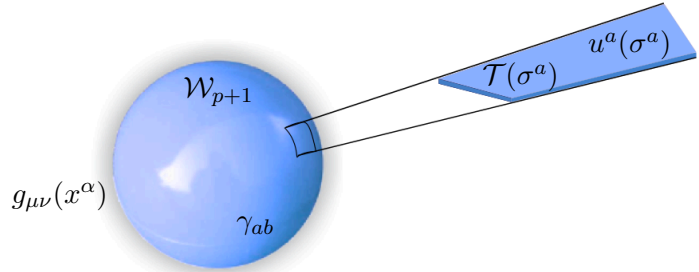
\begin{figure}[!ht]
\centerline{\includegraphics[scale=0.30]{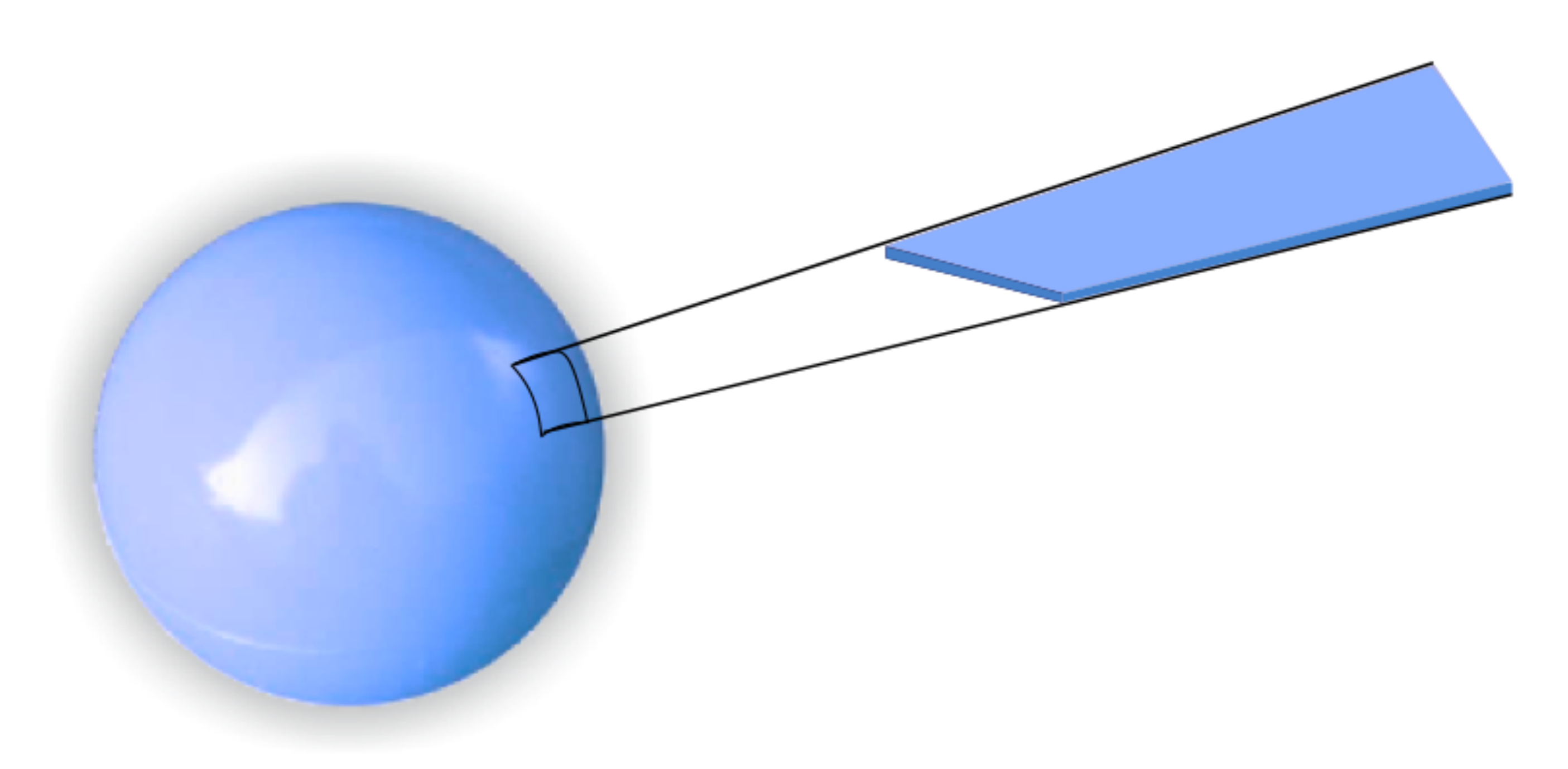} \hskip 1cm}
\begin{picture}(0,0)(0,0)
\put(150,90){ $ \mathcal{W}_{p+1}   $}
\put(170,40){ $ \gamma_{ab}   $}
\put(85,50){ $ g_{\mu\nu}(x^{\alpha})   $}
\put(263,95){ $ \mathcal{T}(\sigma^{a})$}
\put(303,107){ $ u^{a}(\sigma^{a})$}
\end{picture}		
\vskip -1.1cm
\caption{Submanifold embedded in a background space-time locally patched with a perfect fluid.} 
	\label{fig:1}
\end{figure}

We consider submanifolds that span a $(p+1)$-dimensional worldvolume $\mathcal{W}_{p+1}$ embedded in background $D=n+p+3$-dimensional space-time with metric $g_{\mu\nu}(x^{\alpha})$ and coordinates $x^\alpha$ (see Fig.~\ref{fig:1}). The submanifold is parametrized by a set of coordinates $\sigma^a$ and its position in the ambient space-time is parametrized by a set of mapping functions $X^{\mu}(\sigma^{a})$. An arbitrary vector  with support on the worldvolume can be decomposed into tangential and orthogonal components using the respective projectors ${u^{\mu}}_{a}$ and ${n^{\mu}}_{i}$ satisfying ${u^{\mu}}_{a}{n_{\mu}}^{i}=0$, where the indices $a,b,c...$ label worldvolume directions and the indices $i,j,k...$ label transverse directions.

Introducing a complete set of adapted tangential and orthogonal basis in the form $e^{\mu}=\{e^{a}{u^{\mu}}_{a},n^{i}{n^{\mu}}_{i}\}$ we can decompose an arbitrary covector $v_{\mu}$ as $v_{\mu}e^{\mu}=v_{a}e^{a}+v_{i}n^{i}$ where $v_{a}$ and $v_{i}$ are, respectively, the tangent and orthogonal projections of $v_{\mu}$, for example $v_{a}={u^{\mu}}_{a}v_{\mu}$. Given the set of tangential projectors ${u^{\mu}}_{a}\equiv\partial_{a}X^{\mu}$ there is a natural form for the induced metric on the submanifold $\gamma_{ab}\equiv g_{\mu\nu}{u^{\mu}}_{a}{u^{\mu}}_{b}$ where $g_{\mu\nu}$ is evaluated on the surface $x^\alpha=X^{\alpha}(\sigma^{a})$. Since that we will be dealing with tensors with support on the worldvolume, covariant differentiation is only well defined along tangential directions.\footnote{It is possible to have well defined covariant differentiation along orthogonal directions provided we consider a foliation of surfaces \cite{Carter:1997pb}. However we will not consider this here.} Therefore, we introduce the tangential projection of the space-time covariant derivative $\nabla_{a}$ compatible with both the induced and space-time metrics such that acting on an arbitrary tensor field $v^{c\rho}$ reads
\beq
{u^{\mu}}_{a}\nabla_{\mu}v^{c\rho}\equiv\nabla_{a}v^{c\rho}=\partial_{a}v^{c\rho}+{\gamma_{ab}}^{c}v^{b\rho}+\Gamma_{\mu\nu}^{\rho}{u^{\mu}}_{a}v^{c\nu}~~,
\eeq
where ${\gamma_{ab}}^{c}$ are the Christoffel symbols associated with $\gamma_{ab}$ and $\Gamma_{\mu\nu}^{\rho}$ the Christoffel symbols associated with $g_{\mu\nu}$. Given this, the generalization of the classical Gauss-Weingarten equations follows
\beq \label{GW}
\begin{split}
\nabla_{a}\left(e^b{u^\mu}_{b}\right)&={u^{\mu b}}{\gamma_{ab}}^{c}e_c+{n^{\nu}}_{i}{K_{ab}}^{i}e^{a}~~,\\
\nabla_{a}\left(n^{i}{n^\mu}_{i}\right)&=-{u^{\mu b}}{K_{ab}}^{i}n_{i}-{n^\mu}_{j}{\omega_{a}}^{ij}n_{i}~~,
\end{split}
\eeq
where ${K_{ab}}^{i}\equiv{n_{\mu}}^{i}\nabla_{a}{u^{\mu}}_{b}$ is the extrinsic curvature of the embedding, symmetric in its two worldvolume indices $a,b$, and ${\omega_{a}}^{ij}\equiv-{n^\mu}_{j}\nabla_{a}{n_\mu}^{i}$ is the extrinsic twist potential, anti-symmetric in its two transverse indices $i,j$. Therefore, Eqs.\eqref{GW} tell us that the extrinsic curvature is a measure of how the normal basis $n^{i}$ changes along the worldvolume directions while the the extrinsic twist potential tells us how the normals are \emph{twisted} around when displaced in a tangent direction along $\mathcal{W}_{p+1}$. 

It is useful to deal with tangential and orthogonal projections of space-time tensors while still working with space-time indices. For this reason one can introduce the first fundamental form $\gamma^{\mu\nu}\equiv\gamma^{ab}{u^{\mu}}_{a}{u^{\nu}}_{b}$ in order to project along $\mathcal{W}_{p+1}$ and the orthogonal projector $\perp^{\mu\nu}\equiv{n^{\mu i}}{n^{\nu}}_{i}=g^{\mu\nu}-\gamma^{\mu\nu}$, satisfying $\perp^{\mu\nu}\gamma_{\mu\rho}=0$, to project orthogonally to $\mathcal{W}_{p+1}$. Using these structures one can rewrite the second fundamental form as 
\beq \label{extK}
{K_{\mu\nu}}^{\rho}={\gamma^{\lambda}}_{\mu}{\gamma^{\sigma}}_{\nu}\nabla_{\lambda}{\gamma^{\rho}}_{\sigma}=-{\gamma^{\lambda}}_{\mu}{\gamma^{\sigma}}_{\nu}\nabla_{\lambda}{\perp^{\rho}}_{\sigma}~~,
\eeq
which is by definition tangential in its two indices $\mu,\nu$ and orthogonal in the index $\rho$. Using ${u^{\mu}}_{a}$ and ${n^{\mu}}_{i}$ in \eqref{extK} one obtains the extrinsic curvature with worldvolume and transverse indices ${K_{ab}}^{i}={u^{\mu}}_{a}{u^{\nu}}_{b}{n_{\rho}}^{i}{K_{\mu\nu}}^{\rho}$. Furthermore, the tangential projector ${u^{\mu}}_{a}$ is naturally tangential in its space-time index such that ${u^{\mu}}_{a}={\gamma^{\mu}}_{\nu}{u^{\nu}}_{a}$. Similarly, the orthogonal projector ${n^{\mu}}_{i}$ is naturally orthogonal in its space-time index.

The Gauss-Weingarten equations \eqref{GW} do not completely specify the embedded submanifold and must be supplemented by the Gauss-Codazzi, Codazzi-Mainardi and Ricci integrability conditions given by, respectively,
\beq \label{GC}
\begin{split}
R_{abcd}&=\mathcal{R}_{abcd}-{K_{ac}}^{i}K_{bdi}+{K_{ad}}^{i}{K_{bci}}~~,\\
{R^{i}}_{cba}&=\nabla_{b}{{K_{ac}}^{i}}-\nabla_{a}{K_{cb}}^{i}+2{K_{c[a}}^{j}{{\omega_{b]}}^{i}}_{j}~~,\\
{R_{ab}}^{ij}&={\Omega_{ab}}^{ij}-{K_{ac}}^{i}{K_{b}}^{cj}+{K_{bc}}^{i}{K_{a}}^{cj}~~,
\end{split}
\eeq
where we have introduced the Riemann curvature tensor of the background $R_{\mu\nu\lambda\rho}$, the Riemann curvature tensor of the worldvolume $\mathcal{R}_{abcd}$ and the outer curvature tensor associated with the extrinsic twist potential \cite{Capovilla:1994bs},
\beq
{\Omega_{ab}}^{ij}=\nabla_{a}{\omega_{b}}^{ij}-\nabla_{b}\omega_{a}^{ij}+{\omega_{a}}^{ik}{\omega_{bk}}^{j}-{\omega_{b}}^{ik}{\omega_{ak}}^{j}~~.
\eeq
The vanishing of ${\Omega_{ab}}^{ij}$ is the necessary condition for ${\omega_{a}}^{ij}$ to be locally gauged away. For surfaces of codimension-1 both the outer curvature as well as the extrinsic twist potential vanish, as there is only one transverse direction. This completes our review of the geometry of embeddings.


\subsection{Equations of motion} \label{eoms}
The equations of motion satisfied by a space-filling uncharged fluid, ignoring backreaction onto the background, are simply those encompassed by the conservation of the stress-energy tensor $T^{ab}$ associated with the fluid. When confining the fluid to live on an embedded surface the equations of motion, under the same assumptions, are those first obtained by Carter for probe branes \cite{Carter:2000wv} which decompose, respectively, into fluid (intrinsic) and elastic (extrinsic) dynamics as
\beq \label{eqpole}
\nabla_{a}T^{ab}=0~~,~~T^{ab}{K_{ab}}^{i}=0~~.
\eeq  
If one considers corrections to the dynamics of these objects in a derivative expansion, namely the effects of fluctuations of the induced metric, it is necessary to take into account the small, but finite, thickness of the surface itself, and hence expand the stress-energy tensor in a multipole expansion in the manner
\beq \label{pdstress}
T^{\mu\nu}(x^{\alpha})\!=\!\!\int_{\mathcal{W}_{p+1}}\!\!\!\!\!\!\!\!\!d^{p+1}\sigma\sqrt{-\gamma}\left(B^{\mu\nu}(X^{\alpha}(\sigma^a))\frac{\delta^{D}(x^{\alpha}-X^{\alpha})}{\sqrt{-g}}-\nabla_{\rho}\left(B^{\mu\nu\rho}(X^{\alpha}(\sigma^a))\frac{\delta^{D}(x^{\alpha}-X^{\alpha})}{\sqrt{-g}}\right)+...\right)~,
\eeq
where we have slightly generalized the formalism of \cite{Vasilic:2007wp} by allowing the structures $B^{\mu\nu}$ and $B^{\mu\nu\rho}$ to depend on the scalars $X^{\mu}(\sigma^{a})$ instead of just on the worlvolume coordinates $\sigma^{a}$ (see App.~\ref{newform}). These structures introduced above have support on the embedded surface and  can be decomposed as
\begin{eqnarray} \label{bab}\nonumber
B^{ab}&=&T^{ab}+2\mathcal{D}^{(aci}{K^{b)}}_{ci}~~,~~B^{ai}=B^{ia}=-{n^{i}}_{\rho}\nabla_{b}\mathcal{D}^{ab\rho}+\mathcal{S}^{bij}{K^{a}}_{bj}~~,\\
B^{ij}&=& -\mathcal{D}^{ab(i}{K_{ab}}^{j)}~~,~~B^{abi}=-\mathcal{D}^{abi}~~,~~B^{aij}=\mathcal{S}^{aij}~~.
\end{eqnarray}
The tensor structures introduced here can be interpreted in the following way. $T^{ab}$ is the worldvolume stress-energy tensor, $D^{abi}$ is the bending moment of the material and $\mathcal{S}^{aij}$ is the spin current that gives rise to angular momenta in transverse directions to the surface. $T^{ab}$ and $\mathcal{D}^{abi}$ are both symmetric in their worldvolume indices while $\mathcal{S}^{aij}$ is antisymmetric in its transverse indices. 

The equations of motion are then obtained by imposing conservation of the space-time stress-energy tensor 
\beq
\nabla_{\mu} T^{\mu\nu}=0~~,
\eeq
which upon using the methods of \cite{Vasilic:2007wp} can be written in the following way \cite{Armas:2013hsa}
\beq \label{eqdipole1}
\nabla_{a}T^{ab}={u^{b}}_{\mu}\nabla_{a}\nabla_{c}\mathcal{D}^{ac\mu}-\mathcal{D}^{aci}{R^{b}}_{aci}-{\mathcal{S}^a}_{ij}{\Omega_{a}}^{bij}~~,
\eeq
\beq \label{eqdipole2}
T^{ab}{K_{ab}}^{i}={n^{i}}_{\mu}\nabla_a\nabla_b \mathcal{D}^{ab\mu}+\mathcal{D}^{abj}{R^{i}}_{ajb}-2{n^{i}}_{\mu}\nabla_{b}\left({\mathcal{S}_{a}}^{\mu j}{K^{ab}}_{j}\right)+\mathcal{S}^{akj}{R^{i}}_{akj}~~,
\eeq
\beq \label{eqdipole3}
{n_{\mu}}^{i}{n_{\nu}}^{j}\nabla_{a}\mathcal{S}^{a\mu\nu}=-\mathcal{D}^{ab[i}{K_{ab}}^{j]}~~.
\eeq
In the case where the dipole correction $B^{\mu\nu\rho}$ vanishes, and consequently also $\mathcal{D}^{abi}$ and $\mathcal{S}^{aij}$, Eqs.~~\eqref{eqdipole1}-\eqref{eqdipole3} reduce to Eqs.~\eqref{eqpole}. In fact the first two equations above are the modified intrinsic and extrinsic dynamics, respectively, of Eqs.~\eqref{eqpole} while Eq.~\eqref{eqdipole3} is interpreted as a conservation equation for the spin current $\mathcal{S}^{aij}$. 

These equations are invariant under field redefinitions where the position of the surface is displaced by a small amount $\tilde\varepsilon^{i}(\sigma^{a})$ such that $X^{i}(\sigma^{a})\to X^{i}(\sigma^{a})+\tilde\varepsilon^{i}(\sigma^{a})$ while the stress-energy tensor, bending moment and spin current transform as (see App.~\ref{newform})
\beq ~\label{Extra2}
\delta T^{ab}=T^{ab}\tilde\varepsilon^{i}K_{i}-\frac{\partial T^{ab}}{\partial X^{i}}\tilde{\varepsilon}^{i}~~,~~\delta\mathcal{D}^{abi}=T^{ab}\tilde\varepsilon^{i}~~,~~
\delta\mathcal{S}^{aij}=\mathcal{O}\left(\tilde\varepsilon^2\right)~~,
\eeq
where $K^{i}\equiv\gamma^{ab}{K_{ab}}^{i}$ is the mean extrinsic curvature of the embedded surface. This is the usual ambiguity related to the definition of the bending moment $\mathcal{D}^{abi}$ for point particles which in such case can be naturally fixed by choosing the gauge representing the center of mass. For higher dimensional surfaces, there is no natural way of fixing the gauge and hence the bending moment $\mathcal{D}^{abi}$ must be dealt with together with this ambiguity. So far we have been very general and not considered what kind of material $T^{ab}$ represents. Below, we focus in the case of confined fluids and state our assumptions for constructing a theory of dissipative hydrodynamics in a derivative expansion.

\subsection{Confined uncharged and unspinning fluids} \label{confined}
We now wish to apply the equations of motion \eqref{eqdipole1}-\eqref{eqdipole3} to the case of uncharged perfect fluids. For that matter we decompose the stress-energy tensor $T^{ab}$ as 
\beq
T^{ab}=T^{ab}_{(0)}+\Pi^{ab}~~,
\eeq
where $\Pi^{ab}$ denote higher order corrections in the derivative expansion and $T^{ab}_{(0)}$ denotes the perfect fluid stress-energy tensor which we write in the form
\beq \label{stP}
T^{ab}_{(0)}=P\gamma^{ab}+(\epsilon+P)u^{a}u^{b}~~,
\eeq
where $P$ is the fluid pressure, $\epsilon$ its energy density and $u^{a}$ the fluid velocity. Furthermore, one should imagine that each patch of the submanifold where the fluid lives is described by a stress-energy tensor of the form \eqref{stP} to leading order and hence the fluid variables $P,\epsilon,u^{a}, \gamma^{ab}$ are promoted to functions of $X^{\mu}(\sigma^{a})$ on the worldvolune $\mathcal{W}_{p+1}$. The fluid obeys the first law of thermodynamics and the Gibbs-Duhem relations
\beq \label{thermo}
d\epsilon=\mathcal{T}ds~~,~~\epsilon+P=\mathcal{T}s~~,~~dP=sd\mathcal{T}~~,
\eeq
where $s$ and $\mathcal{T}$ denote the local entropy density and temperature of the fluid. Generically, the local thermodynamic fluid variables can be expressed as functions of $\mathcal{T}$, therefore we consider the set of variables $\mathcal{T},u^{a},\gamma^{ab}$ that fully characterize the fluid.\footnote{The local fluid variables vary along the surface. One should see the surface as being locally patched with a perfect fluid to leading order. See Fig~\ref{fig:1}. }

Given this, we now state our assumptions for the construction of the hydrodynamic theory of confined (non)-dissipative fluids:
\begin{itemize}
\item As mentioned above, we assume that the fluid does not backreact onto the background and hence that the equations of motion are those given in \eqref{eqdipole1}-\eqref{eqdipole3}. 

\item We truncate the dissipative theory to first order in the derivative expansion for the case of codimension-1 surfaces and the non-dissipative theory to second order in the derivative expansion in the case of codimension higher than one. In these cases, Eqs.~\eqref{eqdipole1}-\eqref{eqdipole3} capture the full dynamics. In the case of codimension-1 surfaces to second order these equations would have to be modified by including quadrupole corrections \cite{Armas:2013hsa}. Moreover, in order to obtain the right constraints to second order in a dissipative theory it is always necessary to expand the entropy current to third order \cite{Bhattacharyya:2012nq}, which would again require quadrupole corrections. The full form of these equations has not yet been derived in full generality.

\item We assume that the fluid does not carry any spin current, that is, $\mathcal{S}^{aij}=0$. If this was the case, the thermodynamic properties of the fluid \eqref{thermo} would be those analogous to a charged fluid \cite{Armas:2014rva}. While this is an interesting problem, we leave it for future work. In such situations Eqs.~\eqref{eqdipole1}-\eqref{eqdipole2} reduce to
\beq \label{eq1}
\nabla_{a}T^{ab}={n_{\rho}}^{i}{\mathcal{D}^{ac}}_{i}\nabla_{a}{K_{ac}}^{\rho}-2\nabla_{a}\left({\mathcal{D}^{ac}}_{i}{K_{c}}^{bi}\right)~~,
\eeq
\beq \label{eq2}
T^{ab}{K_{ab}}^{i}={n^{i}}_{\mu}\nabla_a\nabla_b \mathcal{D}^{ab\mu}+\mathcal{D}^{abj}{R^{i}}_{ajb}~~,
\eeq
where we have made use of Eq.~\eqref{GC}. Furthemore, Eq.~\eqref{eqdipole3} reduces to the integrability condition
\beq
\mathcal{D}^{ab[i}{K_{ab}}^{j]}=0~~.
\eeq
Note, however, that for codimension-1 surfaces $\mathcal{S}^{aij}$ vanishes anyway, due to the antisymmetry in its two transverse indices and the fact that there is only one transverse index.

\item As a consequence of the last two assumptions, most of the hydrodynamic corrections that $\Pi^{ab}$ can acquire have been classified in \cite{Bhattacharyya:2012nq}, provided one replaces the Riemann curvature tensor considered in \cite{Bhattacharyya:2012nq} by the purely tangential projection of the background Riemann tensor $R_{abcd}$ or the worldvolume Riemann tensor $\mathcal{R}_{abcd}$ since they are related via the Gauss-Codazzi equation given in \eqref{GC}. Therefore the corrections $\Pi^{ab}$ can be decomposed into hydrodynamic and elastic in the form
\beq
\Pi^{ab}=\Pi^{ab}_{\text{hydro}}+\Pi^{ab}_{\text{elastic}}~~.
\eeq
$\Pi^{ab}_{\text{hydro}}$ consist of all the corrections considered in \cite{Bhattacharyya:2012nq} which involve the last worldvolume projection of the background Riemann curvature written in \eqref{GC} while $\Pi^{ab}_{\text{elastic}}$ consists of all the corrections involving ${K_{ab}}^{i}$. Furthermore, as it will be clear below, to the order that we are working, there are no corrections of hydrodynamic nature to the bending moment $\mathcal{D}^{abi}$.

\item We assume a hierarchy of scales between the length scale $R$ associated with the variations of the fluid variables in a neighbourhood of a particular point and the inverse of the local temperature $\mathcal{T}$ at that particular point, namely,
\beq
\frac{1}{\mathcal{T}}\ll R~~.
\eeq
The length scale $R$ is set by the smallest of the scales associated with the mean extrinsic curvature, intrinsic curvature radius or the curvature radius of the background space-time which are typically of the same order according to the Gauss-Codazzi equation \eqref{GC}.

\item We do not consider corrections which are proportional to transverse derivatives of the fluid variables. This is simply because they have been defined as tensor structures with support on the surface as we are not considering a foliation of such surfaces.  We do not account for any corrections proportional to projections of the background Riemann tensor besides those given in Eq.~\eqref{GC}. So far, there are no known examples of fluid configurations with such corrections. Furthermore, we do not consider dimension-dependent corrections, which may be important in the spinning or charged cases \cite{Armas:2013hsa}.

\item Finally, we assume that the fluid is also characterized by a worldvolume entropy current $J^{a}_{s}$ which we require to obey the second law of thermodynamics, that is,
\beq
\nabla_{a}J^{a}_{s}\ge0~~.
\eeq
The requirement of $J^{a}_{s}$ being purely tangential is motivated by two facts. Firstly, there are no known examples where the entropy current acquires transverse components. Secondly, in the case of non-dissipative corrections where $\nabla_{a}J^{a}_{s}=0$ one can show that, on general grounds, requiring an arbitrary space-time current $J^{\mu}(x^{\alpha})$, expanded in a similar manner as in \eqref{pdstress}, to be divergenceless, results in the conservation of a purely tangential worldvolume current \cite{Armas:2012ac, Armas:2013aka, Armas:2014rva}. Furthermore, due to the above assumptions, the divergence of this entropy current can be analyzed independently for the hydrodynamic and elastic corrections to this order. This in fact means that the results obtained in \cite{Bhattacharyya:2012nq} for the corrections there considered still hold in the present case where all quantities should now be treated as worldvolume quantities.
\end{itemize}
Under the assumptions above, in order to construct the theory of dissipative hydrodynamics, it is only necessary to classify the structures appearing in the stress-energy tensor, bending moment and entropy current. In particular, the entropy current can be written as 
\beq
J^{a}_{s}=su^{a}+\mathcal{V}^{a}~~,
\eeq
where $\mathcal{V}^{a}$ includes all the possible higher order corrections. Therefore, we only need to classify all the possible higher order corrections to $\Pi^{ab}, \mathcal{D}^{abi}$ and $\mathcal{V}^{a}$ in terms of derivatives of the fluid variables $\mathcal{T},u^{a},\gamma^{ab}$ and of the background metric $g_{\mu\nu}$. However, note that these fluid variables are not unambiguously defined due to frame transformations and field redefinitions.

\subsubsection*{Frame transformations and field redefinitions}

Under a frame transformation of the form
\beq \label{FR}
\mathcal{T}\to\mathcal{T}+\delta\mathcal{T}~~,~~u^{a}\to u^{a}+\delta u^{a}~~,~~u_{a}\delta u^{a}=0~~,
\eeq
the corrections to the stress-energy tensor $\Pi^{ab}$ and entropy current $\mathcal{V}^{a}$ transform as
\beq \label{FR1}
\Pi^{ab}\to\Pi^{ab}+s\delta\mathcal{T}\gamma^{ab}+\left(s+\mathcal{T}\frac{\partial s}{\partial\mathcal{T}}\right)\delta\mathcal{T}u^{a}u^{b}+2\mathcal{T}su^{(a}\delta u^{b)}~~,
\eeq
\beq\label{FR2}
\mathcal{V}^{a}\to\mathcal{V}^{a}+\frac{\partial s}{\partial \mathcal{T}}\delta\mathcal{T}u^{a}+s\delta u^{a}~~.
\eeq
A standard and convenient choice that we now make is to fix this freedom by choosing the Landau frame, defined as
\beq \label{Landau}
\Pi^{ab}u_{b}=0~~.
\eeq
Note that the field redefinition \eqref{FR} does not affect the bending moment $\mathcal{D}^{abi}$ to the order that we are working. This is because $\mathcal{D}^{abi}$ enters in the equations of motion \eqref{eq1}-\eqref{eq2} by contributing with second or third order terms depending on the surface codimension. Therefore, $\mathcal{D}^{abi}$ can only consist of zeroth order contributions in the case of codimension-1 and of first order contributions in the case of codimension higher than 1. 

Besides the freedom given by the frame transformations \eqref{FR}, there is still the freedom of displacing the embedded surface by a small amount according to \eqref{Extra2}. Defining the transformed bending moment as $\tilde{\mathcal{D}}^{abi}=\mathcal{D}^{abi}+T^{ab}_{(0)}\tilde\varepsilon^{i}$, this freedom can be fixed by different choices of the vector $\tilde\varepsilon^{i}$ that can be obtained by imposing certain constraints, such as
\begin{eqnarray} \nonumber
\textbf{(i)}&&\tilde{\mathcal{D}}^{abi}\gamma_{ab}=0~~,\\ \nonumber
\textbf{(ii)}&&\tilde{\mathcal{D}}^{abi}u_{a}u_{b}=0~~, \\ \nonumber
\textbf{(iii)}&&\tilde{\mathcal{D}}^{abi}P_{ab}=0~~,
\end{eqnarray}
where $P_{ab}=\gamma_{ab}+u_{a}u_{b}$ projects orthogonally to the fluid flows. The most convenient choice which we will consider here is none of the above list but instead we require that no terms proportional to $u^{a}u^{b}K^{i}$ should appear in $\tilde{\mathcal{D}}^{abi}$. One should think of first fixing the choice of surface using the freedom given in \eqref{Extra2} and then imposing the Landau frame condition \eqref{Landau}. Alternatively, we can impose the Landau frame before fixing the choice of surface, in that case the field redefinition $X^{i}(\sigma^{a})\to X^{i}(\sigma^{a})+\tilde\varepsilon^{i}(\sigma^{a})$ yields the  transformation rules written in App.~\ref{newform}. We consider another choice of surface in App.~\ref{basis}.


\subsection{Independent fluid-elastic data} \label{data}
Given the assumptions made in the previous section and the frame choices taken we are now ready to classify all the possible on-shell independent higher order corrections to $\Pi^{ab}, \mathcal{D}^{abi}$ and $J^{a}_{s}$. We classify the necessary new structures to study the case of codimension-1 to first order and of codimension higher than one to second order. We review the  classification scheme for hydrodynamic corrections in App.~\ref{hydro}. For this purpose it is useful to introduce the fluid expansion $\theta$, acceleration $\mathfrak{a}^{a}$, shear $\sigma^{ab}$ and vorticity $\omega^{ab}$ as
\begin{eqnarray} \label{FV} \nonumber
\theta &=&\nabla_{a}u^{a}~~,~~\mathfrak{a}^{a}=u^{b}\nabla_{b}u^{a}~~,\\ 
\sigma^{ab}&=&{P^{ac}}{P^{bd}}\left(\nabla_{(c}u_{d)}-\frac{\theta}{p}\gamma_{cd}\right)~~,~~\omega^{ab}={P^{ac}}{P^{bd}}\nabla_{[c}u_{d]}~~.
\end{eqnarray}
Using this we can decompose the two-tensor $\nabla_{a}u_{b}$ as
\beq
\nabla_{a}u_{b}=-u_{a}\mathfrak{a}_{b}+\sigma_{ab}+\omega_{ab}+\frac{\theta}{p}\gamma_{ab}~~.
\eeq
Our method follows closely \cite{Bhattacharyya:2012nq} but now incorporates other tensor structures which are characteristic of embedded surfaces. The method consists in classifying all possible on-shell independent tensor structures that can appear at a given order in $\Pi^{ab}, \mathcal{D}^{abi}$ and $J^{a}_{s}$. These are constructed from derivatives of the fluid variables  as well as from the extrinsic curvature, which is a first order correction by definition, and the Riemann curvature tensor of the background and of the worldvolume.

In order to classify the independent fluid-elastic data one must use the equations of motion \eqref{eq1}-\eqref{eq2} to exchange certain derivatives by others. For example, to leading order the intrinsic equation of motion \eqref{eqpole} can be projected along parallel and orthogonal directions to the fluid flows, allowing us to express derivatives of the local temperature in terms of the expansion and the acceleration as
\beq
u^{a}\nabla_a\mathcal{T}=-\frac{\partial\mathcal{T}}{\partial s}s\theta~~,~~{P^{ab}}\nabla_{b}\mathcal{T}=-\mathcal{T}\mathfrak{a}^{a}~~.
\eeq
Furthermore, one can always write fluid velocities $u^{a}$ with space-time indices as $u^{\mu}={u^{\mu}}_{a}u^{a}$ and hence the decomposition of \eqref{FV} could be written in terms of space-time indices. However, due to the 
support of these structures on $\mathcal{W}_{p+1}$ only the fluid acceleration can acquire a transverse component as 
$\mathfrak{a}^{i}={n_{\mu}}^{i}u^{a}\nabla_{a}u^{\mu}$. The leading order extrinsic equation \eqref{eqpole} allows us to exchange terms proportional to $\mathfrak{a}^i$, as well as terms of the form $u^{a}u^{b}{K_{ab}}^{i}$ by terms proportional to the mean extrinsic curvature, yielding
\beq
PK^{i}=-(\epsilon+P)u^{a}u^{b}{K_{ab}}^{i}~~,~~PK^{i}=-(\epsilon+P)\mathfrak{a}^{i}~~,
\eeq
where this trivial equality follows from the property that for any purely tangential vector $v^{a}$ we have that $v^{a}v^{b}{K_{ab}}^{i}=\dot{v}^{i}$ with the definition $\dot{v}^{i}={n_{\mu}}^{i}v^{a}\nabla_{a}v^{\mu}$. Using a similar logic for other types of corrections allows us to proceed and classify the possible structures.

Since we will only construct the theory to first order in a derivative expansion in the case of codimension-1 surfaces, the presented analysis will be complete. Therefore, we list the full relevant first order classification of both hydrodynamic and elastic corrections.
\\ 
\renewcommand{\arraystretch}{1.5}
\begin{center} 
    \begin{tabular}{ | c | >{\centering\arraybackslash}m{4.1cm} | c | >{\centering\arraybackslash}m{4.1cm} |}
    \hline
    \color{red}{1st order data} & \color{red}{Before imposing EOM} & \color{red}{EOM} & \color{red}{Independent data} \\ \hline
    Scalars fluid (1) & $u^{a}\nabla_{a}\mathcal{T}~~,~~\theta$ & $u_{b}\nabla_{a}T^{ab}=0$ & $\theta$ \\   \hline
    Vectors fluid (1) & $P^{ab}\nabla_{b}\mathcal{T}~~,~~\mathfrak{a}^{a}$ & ${P^{c}}_{b}\nabla_{a}T^{ab}=0$ & $\mathfrak{a}^{a}$ \\     \hline
    Tensors fluid (1) & $\sigma^{ab}$ & & $\sigma^{ab}$\\     \hline
    Scalars elastic (1) & $\mathfrak{a}^{i}~~,~~K^{i}~~,~~u^{a}u^{b}{K_{ab}}^{i}$ & $T^{ab}{K_{ab}}^{i}=0$ & $K^{i}$   \\ \hline
    Vectors elastic (2) &$u_{b}K^{abi}~~,~~u^{a}K^{i}$& &$u_{b}K^{abi}~~,~~u^{a}K^{i}$\\     \hline 
    Tensors elastic (4) & ${K^{abi}}~~,~~u^{a}u^{b}K^{i}$ $\gamma^{ab}K^{i}~~,~~u^{c}u^{(a}{K_c}^{b)i}$ & & ${K^{abi}}~~,~~u^{a}u^{b}K^{i}$ $\gamma^{ab}K^{i}~~,~~u^{c}u^{(a}{K_c}^{b)i}$  \\     \hline 
     \end{tabular}
         \label{1order} 
\end{center}
\vskip 0.3cm
From the table above we see that there are 7 extra structures that enter the classification when the elastic degrees of freedom are taken into account. Note that we have classified the elastic contributions according to their transformation under worldvolume coordinate transformations. Furthermore, in our classification of tensors we have only considered symmetric tensors in their worldvolume indices, this is because the two tensor structures that we need to classify $T^{ab}$ and $D^{abi}$ are symmetric in their worldvolume indices.

To second order, many more terms of hydrodynamic nature can be added. Since most of them have been classified in \cite{Bhattacharyya:2012nq}, we leave this analysis for App.~\ref{hydro}. Here we list the new terms that can appear due to the presence of the elastic degrees of freedom:
\\ 
\renewcommand{\arraystretch}{1.5}
\begin{center} 
    \begin{tabular}{ | >{\centering\arraybackslash}m{3.2cm} | >{\centering\arraybackslash}m{4.2cm} | c | >{\centering\arraybackslash}m{4.1cm} |}
    \hline
    \color{red}{2nd order data} & \color{red}{Before imposing EOM} & \color{red}{EOM} & \color{red}{Independent data} \\ \hline
    Scalars elastic (3) & $K^{i}K_{i}~~,~~{K^{abi}}{K_{abi}}$ $u^{a}u^{b}{K_{a}}^{ci}{K_{bci}}$ &  & $K^{i}K_{i}~~,~~{K^{abi}}{K_{abi}}$ $u^{a}u^{b}{K_{a}}^{ci}{K_{bci}}$ \\   \hline
    Scalars fluid-elastic (5) & $\theta K^{i}~~,~~\sigma^{ab}{K_{ab}}^{i}$ $\mathfrak{a}^{a}u^{b}{K_{ab}}^{i}~~,~~u^{a}\nabla_{a}K^{\rho}$ $u^{a}\nabla_{b}{K_{a}}^{b\rho}~,~u^{a}u^{b}u^{c}\nabla_{c}{K_{ab}}^{i}$ & $u^{c}\nabla_{c}\left(T^{ab}{K_{ab}}^{\rho}\right)=0$ & $\theta K^{i}~~,~~\sigma^{ab}{K_{ab}}^{i}$ $\mathfrak{a}^{a}u^{b}{K_{ab}}^{i}~~,~~u^{a}\nabla_{a}K^{\rho}$ $u^{a}\nabla_{b}{K_{a}}^{b\rho}$\\     \hline
    Vectors elastic (4) &  $u^{a}K^{i}K_{i}~~,~~u^{a}{K^{bci}}{K_{bci}}$ $u^{a}u^{b}u^{c}{K_{b}}^{di}{K_{cdi}}~~,~~u_{b}{K^{abi}}$  &  & $u^{a}K^{i}K_{i}~~,~~u^{a}{K^{bci}}{K_{bci}}$ $u^{a}u^{b}u^{c}{K_{b}}^{di}{K_{cdi}}~~,~~u_{b}{K^{abi}}$\\     \hline
    Vectors fluid-elastic (11) & $\mathfrak{a}^{a}K^{i}~~,~~\mathfrak{a}_{b}{K^{abi}}$ $\sigma^{ab}u^{c}{K_{bc}}^{i}~~,~~\omega^{ab}u^{c}{K_{bc}}^{i}$ $\nabla^{a}K^{\rho}~~,~~\nabla_{b}{K^{ab\rho}}$ $u^{a}\theta K^{i}~~,~~u^{a}\sigma^{bc}{K_{bc}}^{i}$ $u^{a}\mathfrak{a}^{b}u^{c}{K_{bc}}^{i}~~,~~u^{a}u^{c}\nabla_{c}K^{\rho}$ $\theta u_{b}{K_{a}}^{bi}~,~P^{ad}u^{b}u^{c}\nabla_{d}{K_{bc}}^{i}$ & $P^{ad}\nabla_{d}\left(T^{bc}{K_{bc}}^{\rho}\right)=0$ & $\mathfrak{a}^{a}K^{i}~~,~~\mathfrak{a}_{b}{K^{abi}}$ $\sigma^{ab}u^{c}{K_{bc}}^{i}~~,~~\omega^{ab}u^{c}{K_{bc}}^{i}$ $\nabla^{a}K^{\rho}~~,~~\nabla_{b}{K^{ab\rho}}$ $u^{a}\theta K^{i}~~,~~u^{a}\sigma^{bc}{K_{bc}}^{i}$ $u^{a}\mathfrak{a}^{b}u^{c}{K_{bc}}^{i}~~,~~u^{a}u^{c}\nabla_{c}K^{\rho}$ $\theta u_{b}{K_{a}}^{bi}$   \\ \hline
    Tensors elastic (6) & $K^{abi}K_{i}~~,~~{K^{(a}}_{ci}{K^{b)ci}}$ $u^{c}u^{(a}{K^{b)}}_{ci}K^{i}~~,~~P^{ab}K^{i}K_{i}$ $P^{ab}{K^{cdi}}{K_{cdi}}$ $P^{ab}u^{c}u^{d}{K_{c}}^{ei}{K_{dei}}$ & &$K^{abi}K_{i}~~,~~{K^{(a}}_{ci}{K^{b)ci}}$ $u^{c}u^{(a}{K^{b)}}_{ci}K^{i}~~,~~P^{ab}K^{i}K_{i}$ $P^{ab}{K^{cdi}}{K_{cdi}}$ $P^{ab}u^{c}u^{d}{K_{c}}^{ei}{K_{dei}}$\\     \hline 
     \end{tabular}
         \label{tab:elastic} 
\end{center}
\vskip 0.3cm
A few comments are now in place. In the above table we have only classified the relevant tensors for our purpose. First of all, there are many more tensor structures that could be added to the last row, for example $\gamma^{ab}K^{i}K_{i}$. Moreover, there are many tensors belonging to the category `Tensors fluid-elastic' but these will not be necessary. However, it is necessary to classify third order scalars, as the divergence of a second order quantity - the entropy current - naturally yields third order scalars. The relevant scalars are listed below:
\\ 
\renewcommand{\arraystretch}{1.5}
\begin{center} 
    \begin{tabular}{ | >{\centering\arraybackslash}m{2.7cm}  | >{\centering\arraybackslash}m{4.5cm} | >{\centering\arraybackslash}m{4cm}  | >{\centering\arraybackslash}m{4.6cm} |}
    \hline
    \color{red}{3rd order data} & \color{red}{Before imposing EOM} & \color{red}{EOM} & \color{red}{Independent data} \\ \hline
    Scalars fluid-elastic (16) & $\theta K^{i}K_{i}~~,~~\theta {K^{abi}}{K_{abi}}$ $\theta u^{a}u^{b}{K_{a}}^{ci}{K_{bci}}~~,~~\sigma^{ab}{K_{ab}}^{i}K_{i}$ $\sigma^{ab}{K_{a}}^{ci}{K_{bci}}$ $ \sigma^{ab}u^{c}u^{d}{K_{ac}}^{i}{K_{bdi}}$ $\mathfrak{a}^{a}u^{b}{K_{ab}}^{i}K_{i}~~,~~\mathfrak{a}^{a}u^{b}{K_{a}}^{ci}{K_{bc}}^{i}$ $u^{a}K_{\rho}\nabla_{a}K^{\rho} ~~,~~u^{a}K_{ab\rho}\nabla^{b}K^{\rho}$ $u_{b}K_{\rho}\nabla_{a}K^{ab\rho}$ $u^{a}{K^{bc}}_{\rho}\nabla_{b}{K_{bc}}^{\rho}$ $u_{c}{K^{ab\rho}}\nabla_{a}{K^{c}}_{b\rho}$ $u_{c}{K^{c}}_{b\rho}\nabla_{a}K^{ab\rho}$ $u^{a}u^{b}u^{d}{K_{d}}^{c\rho}\nabla_{b}{K_{ac\rho}}$ $u_{d}u^{c}u^{a}{K^{bd}}_{\rho}\nabla_{a}{K_{cb}}^{\rho}$ $u_{d}u^{c}u^{a}{K^{bd}}_{\rho}\nabla_{b}{K_{ac}}^{\rho}$ $\gamma^{ca}u^{b}K_{\rho}{R^{\rho}}_{cba}~,~u^{b}{K^{ca}}_{\rho}{R^{\rho}}_{cba}$ $u_{d}u^{c}u^{a}{K^{bd}}_{\rho}{R^{\rho}}_{cba}$ & $u^{c}K_{\rho}\nabla_{c}\left(T^{ab}{K_{ab}}^{\rho}\right)=0$ Codazzi-Mainardi Eq.~\eqref{GC} & $\theta K^{i}K_{i}~~,~~\theta {K^{abi}}{K_{abi}}$ $\theta u^{a}u^{b}{K_{a}}^{ci}{K_{bci}}~~,~~\sigma^{ab}{K_{ab}}^{i}K_{i}$ $\sigma^{ab}{K_{a}}^{ci}{K_{bci}}$ $ \sigma^{ab}u^{c}u^{d}{K_{ac}}^{i}{K_{bdi}}$ $\mathfrak{a}^{a}u^{b}{K_{ab}}^{i}K_{i}~~,~~\mathfrak{a}^{a}u^{b}{K_{a}}^{ci}{K_{bc}}^{i}$ $u^{a}K_{\rho}\nabla_{a}K^{\rho} ~~,~~u^{a}K_{ab\rho}\nabla^{b}K^{\rho}$ $u_{b}K_{\rho}\nabla_{a}K^{ab\rho}$ $u^{a}{K^{bc}}_{\rho}\nabla_{b}{K_{bc}}^{\rho}$ $u_{c}{K^{ab\rho}}\nabla_{a}{K^{c}}_{b\rho}$ $u_{c}{K^{c}}_{b\rho}\nabla_{a}K^{ab\rho}$ $u^{a}u^{b}u^{d}{K_{d}}^{c\rho}\nabla_{b}{K_{ac\rho}}$ $u_{d}u^{c}u^{a}{K^{bd}}_{\rho}\nabla_{a}{K_{cb}}^{\rho}$ \\   \hline 
     \end{tabular}
         \label{tab:elastic} 
\end{center}
\vskip 0.3cm
Note that we did not need to classify any structures involving ${\Omega_{ab}}^{ij}$ or ${R^{i}}_{cba}$ introduced in \eqref{GC} since it would require the fluid to be spinning in transverse directions. Moreover note that if the Riemann curvature of the background geometry vanishes then according to the Codazzi-Mainardi equation \eqref{GC} there would be three less independent scalars. For example if the contraction $\gamma^{ca}u^{b}K_{\rho}{R^{\rho}}_{cba}$ vanishes then we have the identity $K_{\rho}u^{a}\nabla_{a}K^{\rho}=K_{\rho}u^{b}\nabla_{a}{K_{b}}^{a\rho}$.

\section{Divergence of the entropy current} \label{divergence}
In this section we compute the divergence of the entropy current to first order in the case of codimension-1 surfaces and to second order in the case of codimension higher than one. The requirement of the second law of thermodynamics to be satisfied imposes constraints on the stress-energy tensor, bending moment and entropy current. We obtain these constraints towards the end of this section.

\subsection{Codimension-1 surfaces} \label{div_co1}
For codimension-1 surfaces and up to first order, our analysis will be fully general and we will describe it here in detail. Using the tables presented in the previous section and App.~\ref{hydro} we can write down the most general stress-energy tensor, bending moment and entropy current as
\beq
T^{ab}=T^{ab}_{(0)}+\eta \sigma^{ab}+\xi \theta P^{ab}+\kappa_1 K P^{ab}+\kappa_2 {P^{a}}_{c}{P^{b}}_{d}K^{cd}~~,
\eeq
\beq
\mathcal{D}^{ab}=\vartheta_1 \gamma^{ab}~~,
\eeq
\beq
J^{a}_{s}=s u^{a}+\beta \theta u^{a}+\gamma \mathfrak{a}^{a}+\pi_{1} K u^{a}+\pi_2 u^{b}{K_{b}}^{a}~~.
\eeq
Note that in the above expressions we have omitted the transverse index since for codimension-1 surfaces there is only one transverse direction. Note that all transport coefficients $\eta,\xi,\kappa_1...$ are functions of the local temperature $\mathcal{T}$. Furthermore, due to the presence of the elastic degrees the freedom, there are 2 extra contributions to the stress-energy tensor and entropy current of the fluid to first order in derivatives.

Calculating the divergence we now find
\beq \label{div1}
\begin{split}
\nabla_{a}J^{a}_{s}=&~\eta\sigma_{ab}\sigma^{ab}+\xi \theta^2\\
&+\theta u^{a}\nabla_{a}\beta +\mathfrak{a}^{a}\nabla_{a}\gamma +\left(\beta+\frac{\gamma}{p}\right)\theta^2+\gamma\left(\omega_{ab}\omega^{ba}+\sigma_{ab}\sigma^{ab}\right)\\
&+\left(\beta+\gamma\right)u^{a}\nabla_a\theta +\gamma u^{a}u^{b}\mathcal{R}_{ab}\\
&+\left(-\frac{\kappa_1}{\mathcal{T}}-2\frac{P}{\mathcal{T}^2}\frac{\partial \vartheta_1}{\partial s}-s\frac{\partial \pi_1}{\partial s}+\pi_1 -\frac{P}{\mathcal{T}}\frac{\partial \pi_2}{\partial s} \right)K\theta\\
&+\frac{1}{p}\left(-\frac{\kappa_2}{\mathcal{T}}+\pi_2\right)P^{ab}K_{ab}+\left(-\mathcal{T}\frac{\partial \pi_2}{\partial\mathcal{T}}-\pi_2-2\frac{\partial \vartheta_1}{\partial \mathcal{T}}\right)u^{a}\mathfrak{a}^{b}K_{ab} +\left(-\frac{\kappa_2}{\mathcal{T}}+\pi_2\right)\sigma^{ab}K_{ab} \\
&+\left(\frac{2\vartheta_1}{\mathcal{T}}+\pi_2\right)u^{b}\nabla_{a}{K^{a}}_{b}+\left(\pi_1-\frac{\vartheta_1}{\mathcal{T}}\right)u^{a}\nabla_{a}K~~.
\end{split}
\eeq
The first three lines of this computation are purely hydrodynamic and have been already computed in \cite{Bhattacharyya:2012nq}. The last three lines are new and constitute the effect of placing the fluid on an embedded surface.

\subsubsection*{Solving for the constraints}
We now require the divergence \eqref{div1} to be positive definite. The procedure for the first three lines is as in \cite{Bhattacharyya:2012nq} which we now review. Since the third line contains terms linear in the fluid data then we must require
\beq
\beta=\gamma=0~~,
\eeq
since otherwise unphysical configurations for which $u^{a}\nabla_{a}\theta$ or $\gamma u^{a}u^{b}\mathcal{R}_{ab}$ are negative would be allowed. This simultaneously eliminates all the terms appearing in the second line. The first line contains only terms which are quadratic in the fluid data, therefore we should only require
\beq \label{etaxi}
\eta\ge 0~~,~~\xi\ge0~~,
\eeq
as previously known in the fluid literature. We now proceed to the analysis of the last three lines. First, we note that all terms appearing in these lines are linear in the fluid data so they must all vanish. The  two terms appearing in the last line are proportional to independent fluid data and hence must be set to zero separately, therefore we must require
\beq \label{c1_1}
\pi_1=\frac{\vartheta_1}{\mathcal{T}}~~,~~\pi_2=-\frac{2\vartheta_1}{\mathcal{T}}~~.
\eeq
The second of these constraints ensures that the second term in the fifth line in \eqref{div1} vanishes. The last term in the fifth line is also independent therefore we must require
\beq\label{c1_2}
\kappa_2=\mathcal{T}\pi_2~~,
\eeq
which ensures that the first term in the fifth line also vanishes. Finally, the term in the fourth line must vanish which therefore requires
\beq\label{c1_3}
\kappa_1=-2 \frac{P}{\mathcal{T}}\frac{\partial \vartheta_1}{\partial s}-\mathcal{T}s\frac{\partial \pi_1}{\partial s}+\mathcal{T}\pi_1 -P\frac{\partial \pi_2}{\partial s}~~.
\eeq
There are three comments worth making about this result. Since $\pi_1$ and $\pi_2$ are expressed in terms of $\vartheta_1$ then so is $\kappa_1$. Therefore, all elastic contributions to the stress-energy tensor and entropy current are uniquely determined in terms of the coefficient $\vartheta_1$ appearing in the bending moment. Furthermore, since all the contributions from these elastic corrections to the divergence \eqref{div1} were required to vanish then such corrections can never be dissipative. This is expected from classical elasticity theory. Moreover, as we have mentioned at the end of Sec.~\ref{data}, if the Riemann curvature tensor of the background geometry vanishes then we have some dependent scalars. In particular the two scalars involved in the last line of \eqref{div1} would be equal to each other. However, the second term in the fifth line of \eqref{div1}, being composed of linear independent data has to vanish and hence requires that both contributions in the last line vanish independently. Finally, we note that terms proportional to $K$ are not invariant under a parity transformation of the normal vector $n^i$. However, since the description of fluid membranes \cite{Helfrich1973, Canham197061, doi:10.1080/00018739700101488} contains such terms we have considered this possibility here.

\subsection{Codimension higher than one} \label{Cod2h}
For codimension higher than one we will be only considering the non-dissipative sector of the theory. We will also only consider here in detail the new terms that appear due to the elastic corrections, since the hydrodynamic corrections have been already considered in \cite{Bhattacharyya:2012nq}. These results however will be reviewed towards the end of Sec.~\ref{discussion}. The most general stress-energy tensor, bending moment and entropy current up to second order can be written as
\beq \label{genst2}
\begin{split}
T^{ab}=&~T^{ab}_{(0)}+\Pi^{ab}_{(1)}+\Pi^{ab}_{(2)}|_{\text{hydro}}+\left(\alpha_1 K^{i}K_{i}+\alpha_2 K^{cdi}K_{cdi}+\alpha_3 u^{c}u^{d}{K_{c}}^{fi}{K_{dfi}}\right)P^{ab} \\
&+{P^{a}}_{c}{P^{b}}_{d}\left(\alpha_4 {K^{cd}}_{i}K^{i}+\alpha_5{K_{f}}^{ci}{K^{fd}}_{i}+\alpha_6 u^{f}u^{h}{K^{c}}_{fi}{K_{h}}^{di}\right)~~,
\end{split}
\eeq 
\beq \label{b2n}
\mathcal{D}^{abi}=\lambda_{1}\gamma^{ab}K^{i}+\lambda_2 K^{abi}+\lambda_3 u^{(a}{K_{c}}^{b)i}u^{c}~~,
\eeq
\beq \label{genE2}
\begin{split}
J^{a}_{s}=&~su^{a}+\mathcal{V}^{a}_{(2)}|_{\text{hydro}}+\left(\beta_1 K^{i}K_{i}+\beta_2 K^{cdi}K_{cdi}+\beta_3 u^{c}u^{d}{K_{c}}^{fi}K_{dfi}\right)u^{a} \\
&+\beta_4 u_{b}K_{i}{K^{abi}}+\beta_5 u_{c}K^{abi}{K^{c}}_{bi}~~.
\end{split}
\eeq
In the above expressions we have introduced $\Pi^{ab}_{(1)}$ which contains the first order corrections to the stress-energy tensor. Since there can be no elastic corrections to first order for codimension higher than one we have that 
\beq
\Pi^{ab}_{(1)}=\eta \sigma^{ab}+\xi\theta P^{ab}~~,
\eeq
where $\eta$ and $\xi$ must satisfy \eqref{etaxi}. We have also introduced $\Pi^{ab}_{(2)}|_{\text{hydro}}$ and $\mathcal{V}^{a}_{(2)}|_{\text{hydro}}$ to denote the hydrodynamic corrections classified in \cite{Bhattacharyya:2012nq}. There are thus 6 additional terms in the stress-energy tensor and 5 additional terms in the entropy current. 

For clarity of presentation we present the divergence of the entropy current for each individual contribution to the bending moment \eqref{b2n} and only taking into account the new elastic corrections since the hydrodynamic ones have been considered in \cite{Bhattacharyya:2012nq} and can be analyzed separately provided neither $\mathcal{R}_{abcd}$ nor $R_{abcd}$ vanish. For the correction corresponding to $\lambda_{1}$ we only need to turn on the contributions that contain $\alpha_1,\alpha_4,\beta_1,\beta_4$, obtaining the divergence
\beq \label{div2_1}
\begin{split}
\nabla_{a}J^{a}_{s}|_{\lambda_1\text{elastic}}=&~\left(-\frac{\alpha_1}{\mathcal{T}}+\beta_1-\frac{2}{\mathcal{T}}\frac{\partial \lambda_1}{\partial s}-s\frac{\partial \beta_1}{\partial s}-\frac{P}{\mathcal{T}}\frac{\partial \beta_4}{\partial s}+\frac{1}{p}\left(-\frac{\alpha_4}{\mathcal{T}}+\beta_{4}\right)\left(1-\frac{P}{\mathcal{T}s}\right)\right)\theta K^{i}K_{i} \\
&+\left(-2\frac{\partial\lambda_1}{\partial\mathcal{T}}-\beta_4-\mathcal{T}\frac{\partial\beta_4}{\partial\mathcal{T}}\right) u^{a}\mathfrak{a}^{b}{K_{ab}}^{i}K_{i}+\left(-\frac{\alpha_4}{\mathcal{T}}+\beta_4\right)\sigma^{ab}{K_{ab}}^{i}K_{i} \\
&+ \left(-\frac{\lambda_1}{\mathcal{T}}+2\beta_1\right)u^{a}K^{\rho}\nabla_{a}K_{\rho}+\left(2\frac{\lambda_1}{\mathcal{T}}+\beta_4\right)\left(u_{b}{K^{ab\rho}}\nabla_{a}K_{\rho}+u_{b}K_{\rho}\nabla_{a}K^{ab\rho}\right)~~.
\end{split}
\eeq

Next, we focus on the contribution coming from $\lambda_2$, in this case we only need to turn on the contributions $\alpha_2,\alpha_5,\beta_2,\beta_5$ and find the divergence
\beq\label{div2_2}
\begin{split}
\nabla_{a}J^{a}_{s}|_{\lambda_2\text{elastic}}=&~\left(-\frac{\alpha_2}{\mathcal{T}}+\beta_2-s\frac{\partial \beta_2}{\partial s}+\frac{1}{p}\left(-\frac{\alpha_5}{\mathcal{T}}+\beta_5\right)\right)\theta K^{abi}K_{abi} \\
&+\left(-\frac{\alpha_3}{\mathcal{T}}+2\frac{s}{\mathcal{T}}\frac{\partial\lambda_2}{\partial s}+s\frac{\partial\beta_5}{\partial s}+\frac{1}{p}\left(-\frac{\alpha_5}{\mathcal{T}}+\beta_5\right)\right)\theta u_a u_{c}{K^{c}}_{bi}K^{abi}\\
&+\left(-2\frac{\partial\lambda_2}{\partial\mathcal{T}}-\beta_5-\mathcal{T}\frac{\partial\beta_5}{\partial\mathcal{T}}\right) u^{a}\mathfrak{a}^{b}{K_{a}}^{ci}K_{bci}+\left(-\frac{\alpha_5}{\mathcal{T}}+\beta_5\right)\sigma^{ab}{K_{a}}^{ci}K_{bci} \\
&+ \left(-\frac{\lambda_2}{\mathcal{T}}+2\beta_2\right)u^{c}K^{ab\rho}\nabla_{c}K_{ab\rho}+\left(2\frac{\lambda_2}{\mathcal{T}}+\beta_5\right)\left({u^{c}K^{ab\rho}}\nabla_{a}K_{cb\rho}+u_{c}K_{bc\rho}\nabla_{a}{K^{a}}_{b\rho}\right)~~.
\end{split}
\eeq

Finally, we consider the contribution from the term proportional to $\lambda_3$ which requires turning on the terms proportional to $\alpha_1,\alpha_3,\alpha_6,\beta_3,\beta_4$. The divergence can be computed as
\beq\label{div2_3}
\begin{split}
\nabla_{a}J^{a}_{s}|_{\lambda_3\text{elastic}}=&~\left(-\frac{\alpha_1}{\mathcal{T}}-\frac{P}{\mathcal{T}}\frac{\partial \beta_4}{\partial s}+P\frac{\lambda_3}{\mathcal{T}^2}\frac{\partial}{\partial s}\left(\frac{P}{\mathcal{T}s}\right)+\frac{s}{\mathcal{T}}\left(\frac{P}{\mathcal{T}s}\right)^2\frac{\partial \lambda_3}{\partial s}\right)\theta K^{i}K_{i} \\
&+\frac{1}{p}\left(-\left(\frac{P}{\mathcal{T}s}\right)^2\frac{\alpha_6}{\mathcal{T}}+\beta_4\left(1-\frac{P}{\mathcal{T}s}\right)-\frac{P}{\mathcal{T}s}\frac{\lambda_3}{\mathcal{T}}\right)\theta K^{i}K_{i} \\
&+\left(-\frac{\alpha_3}{\mathcal{T}}+\beta_3-s\frac{\partial\beta_3}{\partial s}+\frac{\lambda_3}{\mathcal{T}}-\frac{s}{\mathcal{T}}\frac{\partial\lambda_3}{\partial s}-\frac{1}{p\mathcal{T}}\left(\alpha_6+\lambda_3\right)\right)\theta u_a u_{c}{K^{c}}_{bi}K^{abi}\\
&+\left(2\beta_3+\frac{\lambda_3}{\mathcal{T}}\right) u^{a}\mathfrak{a}^{b}{K_{a}}^{ci}K_{bci}\\
& +\left(-\mathcal{T}\frac{\partial \beta_4}{\partial \mathcal{T}}-\beta_4+\frac{\lambda_3}{\mathcal{T}}\frac{\partial}{\partial \mathcal{T}}\left(\frac{P}{\mathcal{T}s}\right)+\frac{P}{\mathcal{T}s}\frac{\partial\lambda_3}{\partial\mathcal{T}}\right) u^{a}\mathfrak{a}^{b}u^{c}u^{d}{K_{ac}}^{i}K_{bdi}\\
&+\left(-\frac{\alpha_6}{\mathcal{T}}-\frac{\lambda_3}{\mathcal{T}}\right)\sigma^{ab}{K_{a}}^{ci}K_{bci}+\left(\beta_4-\frac{P}{\mathcal{T}s}\frac{\lambda_3}{\mathcal{T}}\right)\sigma^{ab}{K_{ab}}^{i}K_{i} \\
&+ \left(2\beta_3+\frac{\lambda_3}{\mathcal{T}}\right)u^{a}u_{b}u^{d}{K_{c}}^{b\rho}\nabla_{a}{K^{c}}_{d\rho}+\left(\beta_4-\frac{P}{\mathcal{T}s}\frac{\lambda_3}{\mathcal{T}}\right)\left(u^{a}{K_{a}}^{b\rho}\nabla_{b}K_{\rho}+u_{b}K_{\rho}\nabla_{a}K^{ab\rho}\right)~.
\end{split}
\eeq
This finalizes the calculations of the divergences. We now proceed and solve for the constraints. 

\subsubsection*{Solving for the constraints}
Since we are interested in the dissipative sector of the theory we impose
\beq\label{saturation}
\nabla_{a}J^{a}_{s}=0~~,
\eeq
which requires all terms appearing in the divergence to vanish. The constraints can be found by imposing \eqref{saturation} for each contribution $\lambda_1,\lambda_2,\lambda_3$ and in the end summing up the individual contributions to each transport coefficient.

We begin by requiring \eqref{div2_1} to vanish. Note that the last line in \eqref{div2_1} is made of independent fluid-elastic data. Therefore one immediately obtains
\beq
\beta_1=\frac{1}{2}\frac{\lambda_1}{\mathcal{T}}~~,~~\beta_4|_{\lambda_1}=-2\frac{\lambda_1}{\mathcal{T}}~~,
\eeq
which in turn leads to the vanishing of the first term in the second line of \eqref{div2_1}. The last term in the second line is also made up of independent fluid-elastic data and hence one must require
\beq
\alpha_4=\mathcal{T}\beta_4~~,
\eeq
which leads to the vanishing of the last term in the first line. Requiring the remaining term in the first line to vanish sets
\beq
\alpha_1|_{\lambda_1}=\mathcal{T}\beta_1-2\frac{P}{\mathcal{T}}\frac{\partial \lambda_1}{\partial s}-\mathcal{T}s\frac{\partial \beta_1}{\partial s}-P\frac{\partial \beta_4}{\partial s}~~.
\eeq

Continuing, we impose the vanishing of \eqref{div2_2}. We first note that the last line in \eqref{div2_2} is constituted by independent fluid-elastic data, therefore we have that
\beq
\beta_2=\frac{1}{2}\frac{\lambda_2}{\mathcal{T}}~~,~~\beta_5=-2\frac{\lambda_2}{\mathcal{T}}~~,
\eeq
which leads to the vanishing of the first term in the third line in \eqref{div2_2}. The last term in the third line is also composed of independent data, hence
\beq
\alpha_5=\mathcal{T}\beta_5~~,
\eeq
leading to the vanishing of the last term in the first and second lines. The last two remaining terms are required to vanish as well and thus we obtain
\beq
\alpha_2=\mathcal{T}\beta_2-\mathcal{T}s\frac{\partial\beta_2}{\partial s}~~,~~\alpha_3|_{\lambda_2}=2s\frac{\partial\lambda_2}{\partial s}+\mathcal{T}s\frac{\partial\beta_5}{\partial s}~~.
\eeq

Lastly, we impose the vanishing of \eqref{div2_3}. The last line in \eqref{div2_3} being composed of independent data leads to the constraints
\beq
\beta_3=-\frac{1}{2}\frac{\lambda_3}{\mathcal{T}}~~,~~\beta_4|_{\lambda_3}=\frac{P}{\mathcal{T}s}\frac{\lambda_3}{\mathcal{T}}~~,
\eeq
which leads to the vanishing of the fourth line, fifth line and of the second term in the sixth line. The first term on the sixth line, being composed of independent data is required to vanish, yielding
\beq
\alpha_6=-\lambda_3~~,
\eeq
leading to the vanishing of the last term in the third line. For the remaining terms we find
\beq
\alpha_1|_{\lambda_3}=-P\frac{\partial \beta_4}{\partial s}+P\frac{\lambda_3}{\mathcal{T}}\frac{\partial}{\partial s}\left(\frac{P}{\mathcal{T}s}\right)+s\left(\frac{P}{\mathcal{T}s}\right)^2\frac{\partial \lambda_3}{\partial s}~~,
\eeq
\beq
\alpha_3|_{\lambda_3}=\mathcal{T}\beta_3-\mathcal{T}s\frac{\partial\beta_3}{\partial s}+\lambda_3-s \frac{\partial\lambda_3}{\partial s}~~.
\eeq

The full system is solved provided one adds up the individual contributions to $\alpha_1,\alpha_3$ and $\beta_4$ such that $\alpha_1=\alpha_1|_{\lambda_1}+\alpha_1|_{\lambda_3}$ without summing over repeated terms and similarly for the other two transport coefficients. We thus obtain the final solution
\beq
\alpha_1=\mathcal{T}\beta_1+2\frac{\partial \lambda_1}{\partial s}-\mathcal{T}s\frac{\partial \beta_1}{\partial s}-P\frac{\partial \beta_4}{\partial s}+P\frac{\lambda_3}{\mathcal{T}}\frac{\partial}{\partial s}\left(\frac{P}{\mathcal{T}s}\right)+s\left(\frac{P}{\mathcal{T}s}\right)^2\frac{\partial \lambda_3}{\partial s}~~,
\eeq
\beq
\alpha_3=2s\frac{\partial\lambda_2}{\partial s}+\mathcal{T}s\frac{\partial\beta_5}{\partial s}+\mathcal{T}\beta_3-\mathcal{T}s\frac{\partial\beta_3}{\partial s}+\lambda_3-s\frac{\partial\lambda_3}{\partial s}~~,
\eeq
\beq
\beta_4=-2\frac{\lambda_1}{\mathcal{T}}+\frac{P}{\mathcal{T}s}\frac{\lambda_3}{\mathcal{T}}~~.
\eeq
Again, as in the case of codimension-1 surfaces, all transport coefficients are determined in terms of the coefficients $\lambda_1,\lambda_2,\lambda_3$ appearing in the bending moment. Also, as in the case of codimension-1 surfaces, if the Riemann tensor of the background geometry vanishes then some of the terms involved in the last line of \eqref{div2_1}-\eqref{div2_3} are equal to each to other. However all the terms involving the acceleration $\mathfrak{a}^{b}$ are composed of linearly independent data and requiring them to vanish leads to the vanishing of the individual contributions involving derivatives of the extrinsic curvature. Therefore, the constraints remain unchanged.  We will now show that these constraints are the same as those obtained from equilibrium partition functions.


\section{Comparison with equilibrium partition functions} \label{functions}
In this section we compare the results of the previous section with the analysis of equilibrium partition functions for confined fluids performed in \cite{Armas:2013hsa} and of the corresponding entropy current perfomed in \cite{Armas:2014rva}. We will show that the constraints arising from these analyses matches the ones found in this work via the study of the divergence of the entropy current. 

To understand how the partition function is obtained according to the analysis of \cite{Armas:2013hsa} we begin by considering the Lorentzian action with the form
\beq \label{actg}
I[X^{\mu}]=\int_{\mathcal{W}_{p+1}}d^{p+1}\sigma\mathcal{L}\left(\sqrt{-\gamma},\mathcal{T},\textbf{k}^{a},\gamma_{ab}, \nabla_{a},{K_{ab}}^{i}\right)~~,
\eeq
where $\textbf{k}^{a}$ is the worldvolume Killing vector field with modulus $\textbf{k}=|-\gamma_{ab}\textbf{k}^{a}\textbf{k}^{b}|^{\frac{1}{2}}$, required for stationarity of the fluid configuration. The dependence of $\mathcal{L}$ on $\mathcal{T}$ can be exchanged by the dependence on the ratio $T/\textbf{k}$ where $T$ is the global (constant) temperature of the overall configuration since $T$ is related by a local redshift of the local temperature via
\beq\label{Tt}
T=\textbf{k}\mathcal{T}~~.
\eeq
The equilibrium partition function can be obtained by first Wick rotating \eqref{actg} and then integrating over the time circle with radius $1/T$ obtaing the free energy

\beq \label{freeg}
\mathcal{F}[X^{\mu}]=-\frac{1}{T}\int_{\mathcal{B}_{p}}\mathcal{L}\left(R_0dV_{(p)},\mathcal{T},\textbf{k}^{a},\gamma_{ab}, \nabla_{a},{K_{ab}}^{i}\right)~~,
\eeq
where we have considered the worldvolume geometry $\mathcal{W}_{p+1}=\mathfrak{R}\times\mathcal{B}_{p}$ and embeddings where $\sqrt{-\gamma}=R_0dV_{(p)}$ with $dV_{p}$ being the volume form on $\mathcal{B}_{p}$. The partition function $Z$ is then obtained simply via the relation
\beq
\ln Z[X^{\mu}]=-\mathcal{F}[X^{\mu}]~~.
\eeq

Since we are interested in the constraints that arise from \eqref{freeg} for the stress-energy tensor, bending moment and entropy current, it is useful to write how these are obtained from \eqref{freeg} \cite{Armas:2013hsa, Armas:2014rva}:
\beq \label{sS}
T^{ab}=\frac{2}{\sqrt{-\gamma}}\frac{\delta \mathcal{L}}{\delta\gamma_{ab}}~~,~~{\mathcal{D}^{ab}}_{i}=\frac{1}{\sqrt{-\gamma}}\frac{\delta \mathcal{L}}{\delta{K_{ab}}^{i}}~~,~~S=-\frac{\partial (T\mathcal{F})}{\partial T}~~,
\eeq
where $S$ is the total entropy. The entropy current $J^{a}_s$ can be obtained from $S$ and reads \cite{Armas:2014rva}
\beq\label{Ec}
J_{s}^{a}=\frac{T}{\mathcal{T}}\frac{\partial \mathcal{L}}{\partial T}u^{a}~~.
\eeq
We will now analyze specific cases of the action \eqref{actg} and compare it with the results of the previous section.

\subsection{Codimension-1 surfaces}
For the codimension-1 surfaces we analyze the most general first order action which takes the form \cite{Armas:2013hsa}
\beq \label{action1}
I[X^{\mu}]=\int_{\mathcal{W}_{p+1}}\sqrt{-\gamma}\left(P(T/\textbf{k})+\tilde \vartheta_1(T/\textbf{k})K\right)~~,
\eeq
where we have omitted the transverse index in the extrinsic curvature. Using \eqref{sS} and noting that for this case $\mathcal{L}=\sqrt{-\gamma}\left(P(T/\textbf{k})+\tilde\lambda(T/\textbf{k})K\right)$ we find
\beq \label{st1}
T^{ab}=T^{ab}_{(0)}+\tilde\lambda(T/\textbf{k}) K \gamma^{ab}-\textbf{k}\tilde\lambda'(T/\textbf{k})Ku^{a}u^{b}~~,~~\mathcal{D}^{ab}=\tilde\vartheta_1(T/\textbf{k})\gamma^{ab}~~,
\eeq 
where the $'$ denotes the derivative with respect to $\textbf{k}$ while we have that
\beq
T^{ab}_{(0)}=P(T/\textbf{k})\gamma^{ab}-\textbf{k}P'(T/\textbf{k})u^{a}u^{b}~~,~~u^{a}=\frac{\textbf{k}^{a}}{\textbf{k}}~~.
\eeq
Using \eqref{Ec} we find the entropy current \cite{Armas:2014rva}
\beq \label{j1}
J^{a}_{s}=su^{a}+\frac{\partial\tilde\vartheta_1}{\partial\mathcal{T}}Ku^{a}~~,
\eeq
where we have used \eqref{thermo}, \eqref{Tt} and suppressed the dependence of the transport coefficient on $\mathcal{T}$. The stress-energy tensor \eqref{st1} and entropy current \eqref{j1} are not written in the Landau gauge \eqref{Landau} and so one must use \eqref{FR1}-\eqref{FR2} to set it in that form. We find that the frame transformation
\beq
\delta\mathcal{T}=\frac{1}{\mathcal{T}}\frac{\partial \mathcal{T}}{\partial s}\left(\tilde\vartheta_1-\mathcal{T}\frac{\partial\tilde\vartheta_1}{\partial\mathcal{T}}-2\frac{P}{\mathcal{T}s}\tilde\vartheta_1\right)K~~,~~\delta u^{a}=-2\frac{\tilde\vartheta_1}{\mathcal{T}s}{P^{ac}}u^{d}K_{cd}~~,
\eeq
brings \eqref{st1} and \eqref{j1} to the Landau gauge, such that
\beq \label{r_1}
\begin{split}
\tilde{T}^{ab}=&T^{ab}_{(0)}+\left(\tilde\vartheta_1+\frac{s}{\mathcal{T}}\frac{\partial \mathcal{T}}{\partial s}\tilde\vartheta_1-s\frac{\partial\tilde\vartheta_1}{\partial s}-2\frac{P}{\mathcal{T}^2}\frac{\partial \mathcal{T}}{\partial s}\tilde\vartheta_1\right)KP^{ab}-2\tilde\vartheta_1 {P^{ac}}P^{bd}K_{cd}~~,\\
\mathcal{D}^{ab}&=\tilde\vartheta_1\gamma^{ab}~~,~~\tilde{J}^{a}_s=su^{a}+\frac{\tilde\vartheta_1}{\mathcal{T}}Ku^{a}-2\frac{\tilde\vartheta_1}{\mathcal{T}}u^{b}{K^{a}}_{b}~~.
\end{split}
\eeq
Note that, as mentioned in the previous section, the bending moment does not transform under a frame transformation to this order. Comparing the above results \eqref{r_1} with those of Sec.~\ref{div_co1} in the stationary case for which the contributions proportional to $\theta$ and $\sigma^{ab}$ vanish and using \eqref{c1_1}-\eqref{c1_3} we find exact agreement provided we identify
\beq
\vartheta_1=\tilde\vartheta_1~~.
\eeq
\subsection{Codimension higher than one}
For codimension higher than one the most general action to second order with extrinsic curvature corrections can be written in the form \cite{Armas:2013hsa}
\beq \label{action2}
I[X^{\mu}]=\int_{\mathcal{W}_{p+1}}\sqrt{-\gamma}\left(P+\tilde\lambda_{1}K^{i}K_{i}+\tilde\lambda_{2}K^{abi}K_{abi}+\tilde\lambda_{3}u^{a}u^{b}{K_{a}}^{ci}K_{bci}\right)~~,
\eeq
where all transport coefficients are functions of the ratio $T/\textbf{k}$. The second order contributions to the stress-energy tensor and the bending moment are summarized in the following table:
\\ 
\renewcommand{\arraystretch}{1.5}
\begin{center} 
    \begin{tabular}{ | c | c | c |}
    \hline
    \color{red}{Scalar} & \color{red}{$\Pi^{ab}_{(2)}|_{\text{elastic}}$} & \color{red}{$\mathcal{D}^{abi}$} \\ \hline
    $\tilde{\lambda_1}K^{i}K_{i}$ & $\tilde\lambda_1K^{i}K_{i}\gamma^{ab}-\tilde\lambda_1'\textbf{k}K^{i}K_{i}u^{a}u^{b}-4\tilde\lambda_1{K^{ab}}_{i}{K}^{i}$ & $2\tilde\lambda_1\gamma^{ab}K^{i}$ \\   \hline
    $\tilde\lambda_2K^{cdi}K_{cdi}$ & $\tilde\lambda_2(\textbf{k})K^{cdi}K_{cdi}\gamma^{ab}-\tilde\lambda_2'\textbf{k}K^{cdi}K_{cdi}u^{a}u^{b}-4\tilde\lambda_2{K^{ac}}_{i}{{K^{b}}_{c}}^{i}$ & $2\tilde\lambda_2 K^{abi}$ \\     \hline
    $\tilde\lambda_3u^{a}u^{b}{K_{a}}^{ci}K_{bci}$ & $\tilde\lambda_3u^{c}u^{d}{K_{c}}^{ei}K_{dei}\gamma^{ab}-(\frac{\tilde\lambda_3}{\textbf{k}^2})'\textbf{k}^3u^{c}u^{d}{K_{c}}^{ei}K_{dei}u^{a}u^{b}-2\tilde\lambda_3u^{c}u^{d}{K^{a}}_{ci}{{K^{b}}_{d}}^{i}$ & $2\tilde\lambda_3u^{c}u^{(a}{{K^{b)}}_{c}}^{i}$ \\     \hline
     \end{tabular}
         \label{tab:elastic} 
\end{center}
\vskip 0.3cm
In particular, the bending moment can be written in the form $\mathcal{D}^{abi}=\mathcal{Y}^{abcd}{K_{cd}}^{i}$ where $\mathcal{Y}^{abcd}$ is the Young modulus of the confined fluid and reads
\beq \label{YM}
\mathcal{Y}^{abcd}=2\left(\tilde\lambda_1\gamma^{ab}\gamma^{cd}+\tilde\lambda_2\gamma^{a(c}\gamma^{d)b}+\tilde\lambda_3 u^{(a}\gamma^{b)(c}u^{d)}\right)~~.
\eeq
The entropy current can be obtained as previously and reads \cite{Armas:2014rva}
\beq \label{ent2}
J_{s}^{a}=su^{a}+\left(\frac{\partial\tilde\lambda_1}{\partial\mathcal{T}}K^{i}K_{i}+\frac{\partial\tilde\lambda_2}{\partial\mathcal{T}}{K^{bci}}K_{bci}+\left(\frac{\partial}{\partial\mathcal{T}}\left(\frac{\tilde\lambda_3}{\textbf{k}^2}\right)-2\frac{\tilde\lambda_3}{\mathcal{T}}\right)u^{b}u^{c}{K_{b}}^{di}{K_{cdi}}\right)u^{a}~~.
\eeq
The stress-energy tensor and the entropy current are not in the Landau gauge. In order to do so we perform a frame transformation with the parameters
\beq
\begin{split}
\delta \mathcal{T}=&~\frac{1}{\mathcal{T}}\frac{\partial\mathcal{T}}{\partial s}\left(\left(\tilde\lambda_1-\mathcal{T}\frac{\partial\tilde\lambda_1}{\partial\mathcal{T}}-4\frac{P}{\mathcal{T}s}\tilde\lambda_1+2\left(\frac{P}{\mathcal{T}s}\right)^2\tilde\lambda_3\right)K^{i}K_{i} +\left(\tilde\lambda_2-\mathcal{T}\frac{\partial \tilde\lambda_2}{\partial\mathcal{T}}\right)K^{abi}K_{abi}\right)\\
&+\frac{1}{\mathcal{T}}\frac{\partial\mathcal{T}}{\partial s}\left(4\tilde\lambda_2-\tilde\lambda_3-\mathcal{T}\frac{\partial\tilde\lambda_3}{\partial\mathcal{T}}\right)~~,\\
\delta u^{a}=&-\frac{1}{\mathcal{T}s}{P^{a}}_{c}\left(4\tilde\lambda_1 u_{b}{K^{cb}}_{i}K^{i}+4\tilde\lambda_2 u^{b} {K^{cd}}_{i}{K_{bd}}^{i}+2\tilde\lambda_3 u_{b}u^{d}u^{e}{K_{d}}^{ci}{K^{b}}_{ei}\right)~~.
\end{split}
\eeq
This transformation brings the stress-energy tensor given in the table above and entropy current \eqref{ent2} to the form \eqref{genst2}-\eqref{genE2} with the coefficients
\beq
\begin{split}
\alpha_1=&\tilde\lambda_{1}\left(1+\frac{s}{\mathcal{T}}\frac{\partial \mathcal{T}}{\partial s}-4\frac{P}{\mathcal{T}s}\right) -s\frac{\partial\tilde\lambda_1}{\partial s}+2\frac{P^2}{\mathcal{T}^3 s}\frac{\partial \mathcal{T}}{\partial s}\tilde\lambda_3~~,~~\alpha_2=\tilde\lambda_2\left(1+\frac{s}{\mathcal{T}}\frac{\partial \mathcal{T}}{\partial s}\right) -s\frac{\partial\tilde\lambda_2}{\partial s}~~,\\
\alpha_3=& 4\frac{s}{\mathcal{T}}\frac{\partial \mathcal{T}}{\partial s}\tilde\lambda_2+\tilde\lambda_3\left(1-\frac{s}{\mathcal{T}}\frac{\partial \mathcal{T}}{\partial s}\right)-s\frac{\partial \mathcal{T}}{\partial s}\frac{\partial\tilde\lambda_3}{\partial \mathcal{T}}~~,~~\alpha_4=-4\tilde\lambda_1~~,~~\alpha_5=-4\tilde\lambda_2~~,~~\alpha_6=-2\tilde\lambda_3~~, \\
\beta_{1}=&\frac{\tilde\lambda_1}{\mathcal{T}}~~,~~\beta_2=\frac{\tilde\lambda_2}{\mathcal{T}}~~,~~\beta_{3}=-\frac{\tilde\lambda_3}{\mathcal{T}}~~,~~\beta_4=-\frac{4}{\mathcal{T}}\tilde\lambda_1+2\frac{P}{\mathcal{T}s}\frac{\tilde\lambda_3}{\mathcal{T}}~~,~~\beta_5=-4\frac{\tilde\lambda_2}{\mathcal{T}}~~.
\end{split}
\eeq
The relations between these coefficients are in exact agreement with the results obtained in Sec.~\ref{Cod2h} provided we identify 
\beq
\lambda_1=2\tilde\lambda_1~~,~~\lambda_2=2\tilde\lambda_2~~,~~\lambda_3=2\tilde\lambda_3~~.
\eeq

\section{Discussion} \label{discussion}
In this work, we have constructed the theory of dissipative hydrodynamics to first order in a derivative expansion in the case of codimension-1 surfaces.\footnote{In the case of higher codimension, to first order there are no elastic corrections} In such cases, the most general stress-energy tensor, entropy current and bending moment read
\beq
T^{ab}=T^{ab}_{(0)}+\eta \sigma^{ab}+\xi \theta P^{ab}+\kappa_1 K P^{ab}+\kappa_2 {P^{a}}_{c}{P^{b}}_{d}K^{cd}~~,
\eeq
\beq
J^{a}_{s}=s u^{a}+\pi_{1} K u^{a}+\pi_2 u^{b}{K_{b}}^{a}~~,~~\mathcal{D}^{ab}=\vartheta_1 \gamma^{ab}~~,
\eeq
where the relations between the several transport coefficients are listed in Sec.~\ref{div_co1}. Hence, the stress-energy tensor is characterized by 2 hydrodynamic and 1 elastic independent transport coefficient. The extra transport coefficient associated with the elastic behaviour is non-dissipative in nature and the way it affects the stress-energy tensor and entropy current is fully determined in terms of the bending moment.

For codimension higher than one, we have constructed the theory of non-dissipative hydrodynamics to second order in a derivative expansion. The most general stress-energy tensor using the results of \cite{Bhattacharyya:2012nq} together with those of Sec.~\ref{Cod2h} is given by\footnote{See App.~\ref{hydro} for the definition of $F^{ab}$ and the operation $<ab>$.}
\beq\label{stfull}
\begin{split}
T^{ab}=&~T^{ab}_{(0)}+\eta\sigma^{ab}+\xi\theta P^{ab} \\
&+\mathcal{T}\Big(\gamma_{1}u^{c}\nabla_{c}\sigma^{<ab>}+\gamma_{2}\mathcal{R}^{<ab>}+\gamma_{3}F^{<ab>}+\gamma_{4}\theta \sigma^{ab} \\
&+\gamma_{5}\sigma^{<ac}{\sigma_{c}}^{b>}+\gamma_{6}\sigma^{<ac}{\omega_{c}}^{b>}+\gamma_{7}\omega^{<ac}{\omega_{c}}^{b>}+\gamma_{8}\mathfrak{a}^{<a}\mathfrak{a}^{b>}\Big) \\
&+\mathcal{T}\left(\zeta_1 u^{c}\nabla_{c}\theta+\zeta_2\mathcal{R}+\zeta_3 u^{c}u^{d}\mathcal{R}_{cd}+\zeta_4\theta^2+\zeta_5\sigma_{cd}\sigma^{cd}+\zeta_6\omega_{cd}\omega^{dc}+\zeta_7\mathfrak{a}^{c}\mathfrak{a}_{c}\right)P^{ab}\\
&+\left(\alpha_1 K^{i}K_{i}+\alpha_2 K^{cdi}K_{cdi}+\alpha_3 u^{c}u^{d}{K_{c}}^{fi}{K_{dfi}}\right)P^{ab} \\
&+{P^{a}}_{c}{P^{b}}_{d}\left(\alpha_4 {K^{cd}}_{i}K^{i}+\alpha_5{K_{f}}^{ci}{K^{fd}}_{i}+\alpha_6 u^{f}u^{h}{K^{c}}_{fi}{K_{h}}^{di}\right)~~,
\end{split}
\eeq
while the entropy current to second order reads
\beq\label{jfull}
\begin{split}
J^{a}_{s}=&~su^{a}+2\nabla_{c}\left(\xi_1u^{[a}\nabla^{c]}\mathcal{T}\right)+\nabla_{c}\left(\xi_2\mathcal{T}\omega^{ac}\right)\\
&+\xi_3\left(\mathcal{R}^{ac}-\frac{1}{2}\gamma^{ac}\mathcal{R}\right)u_{c}+\left(\frac{\xi_3}{\mathcal{T}}+\frac{\partial\xi_3}{\partial\mathcal{T}}\right)\left(\theta\nabla^{a}\mathcal{T}-P^{cd}\nabla_{d}u^{a}\nabla_{c}\mathcal{T}\right)\\
&+\left(\xi_4\omega_{cd}\omega^{dc}+\xi_5\theta^2+\xi_6\sigma^{cd}\sigma_{cd}\right)u^{a}+\xi_7\left(\nabla_{c}s\nabla^{c}su^{a}+2s\theta\nabla^{a}s\right)\\
&+\left(\beta_1 K^{i}K_{i}+\beta_2 K^{cdi}K_{cdi}+\beta_3 u^{c}u^{d}{K_{c}}^{fi}K_{dfi}\right)u^{a} \\
&+\beta_4 u_{b}K_{i}{K^{abi}}+\beta_5 u_{c}K^{abi}{K^{c}}_{bi}~~,
\end{split}
\eeq
and the bending moment is given as
\beq
\mathcal{D}^{abi}=\lambda_1\gamma^{ab}K^{i}+\lambda_2K^{abi}+\lambda_3 u^{c}u^{(a}{K_{c}}^{b)i}~~.
\eeq
The relations between the $\lambda_i$ coefficients and the $\beta_i$ and $\alpha_i$ coefficients are those given in Sec.~\ref{Cod2h}. The relations between the remaining coefficients were obtained in \cite{Bhattacharyya:2012nq} and we write them here for completeness. There are two cases: the case of non-dissipative flows with zero viscosities and the case of non-dissipative and stationary flows with non-zero viscosities. 

For the case of zero viscosities the transport coefficients obey the following relations for $p=3$ \cite{Bhattacharya:2012zx} 
\beq \label{trel}
\begin{split}
\eta=&~\xi=0~~,~~\gamma_{1}=\gamma_{3}+2\xi_6~~,~~\gamma_{2}=\xi_3~~,~~\gamma_{3}=\mathcal{T}\frac{\partial \xi_8}{\partial \mathcal{T}}\\
\gamma_{4}+&\zeta_5=~\xi_6-s\frac{\partial \xi_6}{\partial s}-2s\mathcal{T}\frac{\partial s}{\partial \mathcal{T}}\xi_7~~,~~\gamma_{5}=\gamma_{3}~~,~~\gamma_{7}=\gamma_{3}-4\xi_4~~,\\
\gamma_{8}=&-\mathcal{T}^2\frac{\partial^2\xi_8}{\partial \mathcal{T}^2}-\gamma_{3}-2\xi_7\mathcal{T}^2\left(\frac{\partial s}{\partial \mathcal{T}}\right)^2~~,~~\zeta_1=2s\frac{\partial \xi_8}{\partial s}-\frac{2}{3}\gamma_{3}+2\xi_5+2s^2\xi_7-2\mathcal{T}s\frac{\partial s}{\partial \mathcal{T}}\xi_7~~,\\
\zeta_2=&\frac{1}{2}\left(s\frac{\partial \xi_3}{\partial s}-\frac{\xi_3}{3} \right)~~,~~\zeta_3=2\zeta_2-\frac{2}{3}\gamma_{3}-2\mathcal{T}s\frac{\partial s}{\partial\mathcal{T}}\xi_7~~,\\
 \zeta_4=&-s^2\frac{\partial^2\xi_8}{\partial s^2}-\frac{2}{9}\gamma_{3}+\left(\xi_5-s\frac{\partial \xi_5}{\partial s}\right)-s^3\frac{\partial\xi_7}{\partial s}-s^2\xi_7-\frac{2}{3}\mathcal{T}s\frac{\partial s}{\partial\mathcal{T}}\xi_7\\
 \zeta_6=& -2\mathcal{T}s\frac{\partial s}{\partial\mathcal{T}}\xi_7+\left(\frac{s}{\mathcal{T}}\frac{\partial s}{\partial\mathcal{T}}-\frac{2}{3}\right)\gamma_{3}-s\frac{\partial \xi_4}{\partial s}+\left(2\frac{s}{\mathcal{T}}\frac{\partial s}{\partial\mathcal{T}}-\frac{1}{3}\right)\xi_4~~,\\
 \zeta_7=&\mathcal{T}^2s\frac{\partial s}{\partial\mathcal{T}}\frac{\partial\xi_7}{\partial \mathcal{T}}+\left(\frac{\mathcal{T}^2}{3}\left(\frac{\partial s}{\partial\mathcal{T}}\right)^2+4\mathcal{T}s\frac{\partial s}{\partial\mathcal{T}}+2\mathcal{T}^2s\frac{\partial^2 s}{\partial\mathcal{T}^2}\right)\xi_7+\frac{2}{3}\left(\gamma_{3}+\mathcal{T}^2\frac{\partial^2\xi_8}{\partial\mathcal{T}^2}\right)~~,
\end{split}
\eeq
where we have defined $\xi_8=\xi_3/\mathcal{T}+\partial\xi_3/\partial\mathcal{T}$. In this case, 13 of the hydrodynamic transport coefficients appearing in \eqref{stfull} are fixed in terms of 5 transport coefficients appearing in the entropy current \eqref{jfull}. Including the elastic degrees of freedom, such fluids have a total of 7 hydrodynamic and 3 elastic independent transport coefficients. If the Riemann curvature tensor of the background or the worldvolume geometry vanishes, then there will be a total of 4 independent transport coefficients due to Gauss-Codazzi Eq.~\eqref{GC}.

Now, focusing on the case for which the fluid is stationary, i.e., $\theta=\sigma^{ab}=0$, the stress-energy tensor \eqref{stfull} and entropy current \eqref{jfull} become
\beq\label{stfullstat}
\begin{split}
T^{ab}=&~T^{ab}_{(0)}+\mathcal{T}\left(\gamma_{2}\mathcal{R}^{<ab>}+\gamma_{3}F^{<ab>}+\gamma_{7}\omega^{<ac}{\omega_{c}}^{b>}+\gamma_{8}\mathfrak{a}^{<a}\mathfrak{a}^{b>}\right) \\
&+\mathcal{T}\left(\zeta_2\mathcal{R}+\zeta_3 u^{c}u^{d}\mathcal{R}_{cd}+\zeta_6\omega_{cd}\omega^{dc}+\zeta_7\mathfrak{a}^{c}\mathfrak{a}_{c}\right)P^{ab}\\
&+\left(\alpha_1 K^{i}K_{i}+\alpha_2 K^{cdi}K_{cdi}+\alpha_3 u^{c}u^{d}{K_{c}}^{fi}{K_{dfi}}\right)P^{ab} \\
&+{P^{a}}_{c}{P^{b}}_{d}\left(\alpha_4 {K^{cd}}_{i}K^{i}+\alpha_5{K_{f}}^{ci}{K^{fd}}_{i}+\alpha_6 u^{f}u^{h}{K^{c}}_{fi}{K_{h}}^{di}\right)~~,
\end{split}
\eeq
\beq\label{jfullstat}
\begin{split}
J^{a}_{s}=&~su^{a}+2\nabla_{c}\left(\xi_1u^{[a}\nabla^{c]}\mathcal{T}\right)+\nabla_{c}\left(\xi_2\mathcal{T}\omega^{ac}\right)\\
&+\xi_3\left(\mathcal{R}^{ac}-\frac{1}{2}\gamma^{ac}\mathcal{R}\right)u_{c}-\left(\frac{\xi_3}{\mathcal{T}}+\frac{\partial\xi_3}{\partial\mathcal{T}}\right)P^{cd}\nabla_{d}u^{a}\nabla_{c}\mathcal{T}\\
&+\xi_4\omega_{cd}\omega^{dc}u^{a}+\xi_7\nabla_{c}s\nabla^{c}su^{a}\\
&+\left(\beta_1 K^{i}K_{i}+\beta_2 K^{cdi}K_{cdi}+\beta_3 u^{c}u^{d}{K_{c}}^{fi}K_{dfi}\right)u^{a} \\
&+\beta_4 u_{b}K_{i}{K^{abi}}+\beta_5 u_{c}K^{abi}{K^{c}}_{bi}~~.
\end{split}
\eeq
The relations between the hydrodynamic transport coefficients are those presented in \eqref{trel} but now the viscosities are allowed to be $\eta\ge0~,~\xi\ge0$. Such relations for hydrodynamic transport coefficients can also be obtained from equilibrium partition functions as in Sec.~\ref{functions} and this has been done in \cite{Banerjee:2012iz}. The 8 non-dissipative hydrodynamic coefficients appearing in the stress-tensor \eqref{stfullstat} are now fixed in terms of the 3 independent transport coefficients appearing in the entropy current \eqref{jfullstat}. In this case there are 3 hydrodynamic and 3 elastic independent transport coefficients to second order in the expansion when including the elastic degrees of freedom and a total of 4 if the Riemann tensor of the background or the worldvolume geometry vanishes.

When considering the measurement of these transport coefficients from gravity, in particular those associated with the elastic degrees of freedom, the object which can be measured directly is $B^{ab}$ instead of $T^{ab}$, even though one can always be exchanged by the other using relation \eqref{bab}. It is therefore useful to present the results for $B^{ab}$ in the Landau gauge and corresponding entropy current. This is done in the end of App.~\ref{basis}.

We note that we have only constructed the theory to first order for codimension-1 surfaces and the non-dissipative sector to second order for codimension higher than one. This was due to the fact that in order to allow for dissipation to second order it is necessary to obtain the pole-quadrupole equations of motion in the spirit of \cite{Vasilic:2007wp}. However, it is expected from classical elasticity theory that all corrections induced via the bending moment to this order are non-dissipative and hence the results here apply even in the case of dissipative flows where the results of \cite{Bhattacharyya:2012nq} for the hydrodynamic corrections should be taken into account. We leave a precise check of this for future work. 

Finally, allowing for the fluid to be electrically charged or spinning in transverse directions to the surface would provide interesting connections to charged and doubly-spinning black holes. Moreover, perturbing a black brane both intrinsically and extrinsically in a time-dependent setting would allow to observe the different relations between the transport coefficients. Understanding the role of the elastic corrections in an AdS/CFT context would be worthwhile pursuing.

\section*{Acknowledgements}
I am extremely grateful (again) to Jyotirmoy Bhattacharya for the many lessons on hydrodynamics, for many useful discussions and comments on a draft of this paper. I am also thankful to Sayantani Bhattacharyya for useful correspondence. I am indebted to Troels Harmark for many discussions and related work \cite{Armas:2014rva} that allowed me to perform some of the checks in Sec.~\ref{functions}.  I am also thankful to Niels Obers and Joan Camps for comments on an early draft of this paper. I would like to thank the organizers of the workshop \textbf{New perspectives from Gravity in Higher-Dimensions} in Benasque (2013), of the workshop \textbf{Current Themes in High Energy Physics and Cosmology} at NBI (2013) and of \textbf{The String Theory Universe} in Bern (2013) during which parts of this work were carried out. I also acknowledge CERN and NBI for hospitality and Fabrikken/Christiania for support. This work has been supported by the Swiss National Science Foundation and the `Innovations- und Kooperationsprojekt C-13' of the Schweizerische Universit\"{a}tskonferenz SUK/CUS.

\appendix

\section{Generalization of the multipole expansion formalism} \label{newform}
As mentioned in Sec.~\ref{eoms}, in order to work with fluids living on embedded surfaces it is necessary to make a slight extension of the formalism developed in \cite{Vasilic:2007wp}, in particular the tensor structures $B^{\mu\nu}$ and $B^{\mu\nu\rho}$, as well as other structures appearing at higher orders in the expansion, should be allowed to depend on the set of mapping functions $X^{\mu}(\sigma^{a})$ instead of just on the worldvolume coordinates $\sigma^{a}$. This generalization does not modify the results of \cite{Vasilic:2007wp} except for the transformations of these tensor structures under the field redefinition\footnote{This field redefinition was coined by the authors of \cite{Vasilic:2007wp} as `extra symmetry 2'.}
\beq \label{field}
X^{\mu}(\sigma^{a})\to \bar{X}^{\mu}(\sigma^{a})=X^{\mu}(\sigma^{a})+\tilde\varepsilon^{\mu}(\sigma^{a})~~,
\eeq
where $\bar{X}(\sigma^{a})$ represents the set of mapping functions describing the position of the new surface. Under this transformation the induced metric on the surface as well as the stress-energy tensor components change to $\bar{\gamma}_{ab}~,~\bar{B}^{\mu\nu}(\bar{X}^{u}(\sigma^{a}))~,~\bar{B}^{\mu\nu\rho}(\bar{X}^{u}(\sigma^{a}))$. Using \eqref{field}, we find that the stress-energy tensor \eqref{pdstress} transforms to
\beq \label{pdstress}
T^{\mu\nu}(x^{\alpha})\!=\!\int_{\bar{\mathcal{W}}_{p+1}}\!\!\!\!\!\!\!d^{p+1}\sigma\sqrt{-\bar\gamma}\left(\bar B^{\mu\nu}(\bar X(\sigma^{a}))\frac{\delta^{D}(x^{\alpha}-\bar X^{\alpha}(\sigma^a))}{\sqrt{-g}}-\nabla_{\rho}\left(\bar B^{\mu\nu\rho}(\bar{X}(\sigma^{a}))\frac{\delta^{D}(x^{\alpha}-\bar X^{\alpha}(\sigma^a))}{\sqrt{-g}}\right)+...\right)~,
\eeq
where the new tensor structures differ from the old ones by \footnote{Note that in deriving these transformation rules, we have used that $\partial_{\rho}B^{\mu\nu}(X^{\alpha}(\sigma^{a}))=0$ since $B^{\mu\nu}$ is evaluated on the surface $X^{\alpha}(\sigma^{a})$.}
\beq \label{extra2_2}
\delta B^{\mu\nu}=-B^{\mu\nu}{u^{a}}_{\rho}\nabla_{a}\tilde \varepsilon^{\rho}-2\tilde{\varepsilon}^{\rho}{\Gamma^{(\mu}}_{\lambda\rho}B^{\nu)\lambda}-{E^{\mu\nu}}_{\rho}~\tilde\varepsilon^{\rho}~,~~\delta B^{\mu\nu\rho}=-B^{\mu\nu}\tilde\varepsilon^{\rho}~~,
\eeq
where we have defined the tensor ${E^{\mu\nu}}_{\rho}$ symmetric in the indices $\mu,\nu$ as 
\beq
{E^{\mu\nu}}_{\rho}=\frac{\partial B^{\mu\nu}}{\partial X^{\rho}}~~.
\eeq
The appearance of ${E^{\mu\nu}}_{\rho}$ in Eq.~\eqref{extra2_2} is the main difference from the work of \cite{Vasilic:2007wp}. Under these transformation rules, the equations of motion presented in Sec.~\ref{eoms} remain invariant. This implies that the worldvolume stress-energy tensor $T^{ab}$ and the bending moment $\mathcal{D}^{abi}$ transform as
\beq \label{extra2_3}
\delta T^{ab}=T^{ab}\tilde\varepsilon^{i}K_{i}+\left(\tilde{\varepsilon}^{c}\nabla_{c}T^{ab}-2T^{c(a}\nabla_{c}\tilde{\varepsilon}^{b)}\right) -{E^{ab}}_{\rho}\tilde\varepsilon^{\rho}~~,~~\delta\mathcal{D}^{abi}=T^{ab}\tilde{\varepsilon}^{i}~~,
\eeq
where we have decomposed the deformation vector as $\tilde{\varepsilon}^{\mu}=\tilde{\varepsilon}^{a}{u^{\mu}}_{a}+\tilde{\varepsilon}^{i}{n^{\mu}}_{i}$. The variation of the spin current is of higher order.

These transformation rules can also be seen from the equations of motion. The field redefinition \eqref{field} is only non-trivial along transverse directions as it coincides with worldvolume reparametrizations along wordvolume directions\footnote{This is true except at the boundary \cite{Vasilic:2007wp}.}, so it suffices to analyze the extrinsic equation of motion \eqref{eq2} evaluated on the surface $\bar{X}(\sigma^{a})$, 
\beq \label{eomS1}
\int_{\bar{\mathcal{W}}_{p+1}}d^{p+1}\sigma\sqrt{-\bar\gamma}\left(\bar{T}^{ab}{\bar{K}_{ab}}^{i}-{n^{i}}_{\mu}\nabla_a\nabla_b \bar{\mathcal{D}}^{ab\mu}-\bar{\mathcal{D}}^{abj}{\bar{R}^{i}}_{ajb}\right)=0~~.
\eeq
Here we have not taken into account the spin current since it does not transform under field redefinitions to this order. Here we have also written the equations of motion with the integration over the surface since this is the form in which they are obtained \cite{Vasilic:2007wp}. Including the integration does not modify the equation of motion but it is necessary in order to obtain the correct transformation properties under \eqref{field}. We now write Eq.~\eqref{eomS1} in terms of quantities evaluated on the surface $X^{\mu}(\sigma^{a})$ using the transformation rules along transverse directions for the induced metric and extrinsic curvatures \cite{Armas:2013hsa}
\beq
\delta\sqrt{-\gamma}=-\sqrt{-\gamma}\tilde{\varepsilon}^{i}K_{i}~~,~~\delta{K_{ab}}^{i}={n^{i}}_{\mu}\nabla_{a}\nabla_{b}\tilde{\varepsilon}^{\mu}-{R^{i}}_{baj}\tilde{\varepsilon}^{j}~~,
\eeq
bringing \eqref{eomS1} to the form
\beq \label{eomS2}
\int_{\mathcal{W}_{p+1}}\!\!\!\!\! d^{p+1}\sigma\sqrt{-\gamma}\left((\bar{T}^{ab}-T^{ab}\tilde{\varepsilon}^{i}K_{i}+{E^{ab}}_{i}\tilde{\varepsilon}^{i}){K_{ab}}^{i}-{n^{i}}_{\mu}\nabla_a\nabla_b (\bar{\mathcal{D}}^{ab\mu}-T^{ab}\tilde{\varepsilon}^{\mu})-(\bar{\mathcal{D}}^{abj}-T^{ab}\tilde{\varepsilon}^{j}){\bar{R}^{i}}_{ajb}\right)=0~~.
\eeq
We see that the stress-energy tensor and the bending moment have transformed in the opposite way as \eqref{extra2_3} and hence yield the equation of motion on the surface $X^{\mu}(\sigma^{a})$.

\subsection*{Frame transformations under field redefinitions}
Here we analyze the transformation properties in the case for which $T^{ab}$ is of the perfect fluid form to leading order. In this case, the field redefinition \eqref{extra2_2}, besides introducing new contributions to the stress-energy tensor, also induces a frame transformation in the fluid variables. According to \eqref{extra2_3} the variation of the stress-energy tensor is
\beq \label{varst}
\delta T^{ab}=T^{ab}_{(0)}\tilde{\varepsilon}^{i}K_{i}-\left(\frac{\partial P}{\partial X^{i}}\gamma^{ab}+\frac{\partial (\epsilon + P)}{\partial X^{i}}u^{a}u^{b}+2(\epsilon+P)\frac{\partial u^{(a}}{\partial X^{i}}u^{b)}\right)\tilde{\varepsilon}^{i} -P {K^{ab}}_{i}\tilde{\varepsilon}^{i}~~.
\eeq
Indeed the middle term above can be interpreted as a frame transformation with parameters
\beq \label{frame1}
\delta \mathcal{T}=-\frac{\partial\mathcal{T}}{\partial X^{i}}\tilde{\varepsilon}^{i}~~,~~\delta u^{a}=-\frac{\partial u^{a}}{\partial X^{i}}\tilde{\varepsilon}^{i}~~.
\eeq
Furthermore, in the case of stationary fluids \cite{Armas:2013hsa}, the fluid variables depend on the scalars $X^{i}(\sigma)$ only via the induced metric. In this case the variation \eqref{varst} can be rewritten as 
\beq \label{varst2}
\delta T^{ab}=T^{ab}_{(0)}\tilde{\varepsilon}^{i}K_{i}-E^{abcd}{K_{cd}}^{i}\tilde{\varepsilon}_{i}~~,
\eeq
where $E^{abcd}$ is the elasticity tensor of the confined fluid to leading order defined as \cite{Armas:2012jg}
\beq \label{elasticity}
E^{abcd}=-2\frac{\partial T^{ab}_{(0)}}{\partial\gamma_{cd}}~~.
\eeq
The transformation properties of the entropy current can be obtained by generalizing the analysis of \cite{Armas:2012ac, Armas:2013aka} and promoting the worldvolume functions to functions of the scalars $X^{\mu}(\sigma)$. From this generalization presented in \cite{Armas:2014rva}, one finds the transformation rule for the fluid entropy current,
\beq
\delta J^{a}_{s}=su^{a}\tilde{\varepsilon}^{i}K_{i}-\left(\frac{\partial s}{\partial X^{i}}+s\frac{\partial u^{a}}{\partial X^{i}}\right)\tilde{\varepsilon}^{i}~~.
\eeq
Again, we can see this variation as a frame transformation with parameters \eqref{frame1}. As mentioned in Sec.~\ref{confined}, one can present these transformations in Landau gauge. In order to do so we perform the frame transformation with parameters
\beq
\delta \mathcal{T}=-\frac{1}{\mathcal{T}}\frac{\partial\mathcal{T}}{\partial s}\left(\epsilon+\frac{P^2}{\mathcal{T}s}\right)\tilde{\varepsilon}^{i}K_{i}+\frac{\partial\mathcal{T}}{\partial X^{i}}\tilde{\varepsilon}^{i}~~,~~\delta u^{a}=-\frac{P}{\mathcal{T}s}{P^{a}}_{c}u_{d}{K^{cd}}_{i}\tilde{\varepsilon}^{i}+\frac{\partial u^{a}}{\partial X^{i}}\tilde{\varepsilon}^{i}~~,
\eeq
bringing the variations to the form
\beq 
\begin{split}
\delta T^{ab}=&\left(P-s\frac{\epsilon}{\mathcal{T}}\frac{\partial \mathcal{T}}{\partial s}-\frac{P^2}{\mathcal{T}^2}\frac{\partial \mathcal{T}}{\partial s}\right) P^{ab}\tilde\varepsilon^{i}K_{i}-P{P^{a}}_{c}{P^{b}}_{d}{K^{cd}}_{i}\tilde\varepsilon^{i}~~,~~\delta\mathcal{D}^{abi}=T^{ab}_{(0)}\tilde\varepsilon^{i}~~,\\
\delta J^{a}_{s}=&-\frac{\epsilon}{\mathcal{T}}u^{a}\tilde\varepsilon^{i}K_{i}-\frac{P}{\mathcal{T}}u_{b}{K^{ab}}_{i}\tilde{\varepsilon}^{i}~~.
\end{split}
\eeq

\section{Review of independent fluid data} \label{hydro}
Here we review the fluid data classified in \cite{Bhattacharyya:2012nq} which is necessary for obtaining the results of Sec.~\ref{div_co1} and Sec.~\ref{discussion} for the hydrodynamic corrections.  

To first order in the expansion we have already classified the relevant independent fluid data in Sec.~\ref{data}. To second order we have the following independent data:
\\ 
\renewcommand{\arraystretch}{1.5}
\begin{center} 
    \begin{tabular}{ | c | >{\centering\arraybackslash}m{4.1cm} | >{\centering\arraybackslash}m{4.1cm} | >{\centering\arraybackslash}m{4.1cm} |}
    \hline
    \color{red}{2nd order data} & \color{red}{Before imposing EOM} & \color{red}{EOM} & \color{red}{Independent data} \\ \hline
    Scalars fluid (3) & $u^{a}\nabla_{a}\theta~~,~~~\nabla_{a}\nabla^{a}\mathcal{T}$ $u^{a}u^{b}\nabla_{a}\nabla_{b}\mathcal{T}~~,~~\mathcal{R}$ $\gamma^{ac}\gamma^{bd}R_{abcd}~~,~~u^{a}u^{b}\mathcal{R}_{ab}$ $u^{a}u^{b}\gamma^{cd}R_{acbd}$& $u_{b}u^{a}\nabla_{a}\nabla_{c}T^{bc}=0$ $\nabla_{a}\nabla_{b}T^{ab}=0$ Gauss-Codazzi Eq.~\eqref{GC} &$u^{a}\nabla_{a}\theta~~,~~\mathcal{R}~~,~~u^{a}u^{b}\mathcal{R}_{ab}$  \\   \hline
    Vectors fluid (3) & $P^{ab}u^{c}\nabla_{c}\mathfrak{a}_{b}~~,~~P^{ab}\nabla_{c}\nabla^{c}u_{b}$ $P^{ab}\nabla_{b}\theta~~,~P^{ab}\nabla_{b}(u^{c}\nabla_{c})\mathcal{T}$ $P^{ab}\mathcal{R}_{bc}u^{c}~,~P^{ab}\gamma^{de}R_{dbec}u^{c}$ & ${P^{a}}_{b}u^{c}\nabla_{c}\nabla_{d}T^{db}=0$ $u_{c}P^{ab}\nabla_{b}\nabla_{d}T^{cd}=0$  Gauss-Codazzi Eq.~\eqref{GC}&  $P^{ab}\nabla_{b}\theta~~,~{P^{a}}_{b}\nabla_{c}\sigma^{bc}$ $P^{ab}\mathcal{R}_{bc}u^{c}$ \\     \hline
    Tensors fluid (3) & ${P_{a}}^{c}{P_{b}}^{d}u^{e}\nabla_{e}\sigma_{cd}$ $\nabla_{<a}\nabla_{b>}\mathcal{T}$ $\mathcal{R}_{<ab>}~~,~~F_{<ab>}$ $\gamma^{cd}R_{c<adb>}~,~u^{c}u^{d}R_{<acb>d}$ & $\nabla_{<a}\nabla_{c}T^{c}_{b>}=0$ Gauss-Codazzi Eq.~\eqref{GC} & ${P_{a}}^{c}{P_{b}}^{d}u^{e}\nabla_{e}\sigma_{cd}$ $\mathcal{R}_{<ab>}~~,~~F_{<ab>}$\\     \hline
    Spin-3 (1) &$\nabla_{<a}\nabla_{b}u_{c>}$& &$\nabla_{<a}\nabla_{b}u_{c>}$\\     \hline 
     \end{tabular}
         \label{1order} 
\end{center}
\vskip 0.3cm
In the above list we have defined the tensor $F^{ab}$
\beq
F_{ab}=u^{c}u^{d}\mathcal{R}_{acbd}~~,
\eeq
as well as the operation $<ab>$ on any two-tensor $A_{ab}$ as
\beq
A_{<ab>}={P^{c}}_{a}{P^{d}}_{b}\left(\frac{A_{cd}+A_{dc}}{2}-\gamma_{cd}\frac{P^{ef}A_{ef}}{p}\right)~~.
\eeq
To second order we also have the composite independent data:
\\ 
\renewcommand{\arraystretch}{1.5}
\begin{center} 
    \begin{tabular}{ | c |  >{\centering\arraybackslash}m{10cm} |}
    \hline
    \color{red}{2nd order data} & \color{red}{Independent composite data} \\ \hline
    Scalars fluid (4) &  $\theta^2~~,~~\mathfrak{a}^{c}\mathfrak{a}_{c}~~,~~\sigma_{ab}\sigma^{ab}~~,~~\omega_{ab}\omega^{ba}$  \\   \hline
    Vectors fluid (3) & $\mathfrak{a}^{a}\theta~~,~~\mathfrak{a}_{b}\omega^{ab}~~,~~\mathfrak{a}_{b}\sigma^{ab}$  \\     \hline
    Tensors fluid (5) & $\theta\sigma_{ab}~~,~~{\sigma^{c}}_{<a}\sigma_{cb>}~~,~~{\omega^{c}}_{<a}\sigma_{cb>}~~,~~{\omega^{c}}_{<a}\omega_{cb>}~~,~~\mathfrak{a}_{<a}\mathfrak{a}_{b>}$  \\     \hline
     \end{tabular}
         \label{1order} 
\end{center}
\vskip 0.3cm
To third order we only need to classify the relevant scalars. These are (for $p=3$):
\\ 
\renewcommand{\arraystretch}{1.5}
\begin{center} 
    \begin{tabular}{ | c | >{\centering\arraybackslash}m{4.1cm} | >{\centering\arraybackslash}m{4.2cm} | >{\centering\arraybackslash}m{4.1cm} |}
    \hline
    \color{red}{3rd order data} & \color{red}{Before imposing EOM} & \color{red}{EOM} & \color{red}{Independent data} \\ \hline
    Scalars fluid (3) & $u^{c}\nabla_{c}(u^{d}\nabla_{d}\theta)~~,~~\nabla_{c}\nabla^{c}\theta$ $u^{c}\nabla_{c}(u^{d}\nabla_{d}(u^{b}\nabla_{b}\mathcal{T}))$ $u^{c}\nabla_{c}\nabla_{d}\nabla^{d}\mathcal{T}$ $u^{c}\nabla_{c}\mathcal{R}~~,~~u^{c}\nabla_{c}(u^{a}u^{b}\mathcal{R}_{ab})$ $u_{a}\nabla_{b}\mathcal{R}^{ab}$ $u^{e}\nabla_{e}(\gamma^{ac}\gamma^{bd}R_{abcd})$ $u^{c}\nabla_{c}(u^{a}u^{b}\gamma^{de}R_{daeb})$ $u_{a}\nabla_{b}(\gamma_{cd}R^{cadb})$ & $u_{c}u^{d}\nabla_{d}(u^{e}\nabla_{e}\nabla_{b}T^{bc})=0$ $u^{c}\nabla_{c}\nabla_{a}\nabla_{b}T^{ab}=0$ $u_{c}\nabla_{d}\nabla^{d}\nabla_{b}T^{bc}=0$ $u^{c}\epsilon_{cabd}\epsilon^{aefg}\nabla_{e}{\mathcal{R}^{bd}}_{fg}=0$ Gauss-Codazzi Eq.~\eqref{GC} &$u^{c}\nabla_{c}(u^{d}\nabla_{d}\theta)$  $u^{c}\nabla_{c}\mathcal{R}~~,~~u^{c}\nabla_{c}(u^{a}u^{b}\mathcal{R}_{ab})$  \\   \hline
     \end{tabular}
         \label{1order} 
\end{center}
\vskip 0.3cm
Finally, we also have independent composite scalars to third order in the expansion, which areß listed below:
\renewcommand{\arraystretch}{1.5}
\begin{center} 
    \begin{tabular}{ | c |  >{\centering\arraybackslash}m{14cm} |}
    \hline
    \color{red}{3rd order data} & \color{red}{Independent composite data} \\ \hline
    Scalars fluid (16) &  $\theta u^{c}\nabla_{c}\theta~~,~~(\nabla_{c}\mathcal{T})\nabla_{d}\nabla_{d}u^{c}~~,~~(\nabla_{c}\mathcal{T})u^{d}\nabla_{d}\nabla^{c}\mathcal{T}~~,~~\sigma_{ab}\nabla^{a}\nabla^{b}\mathcal{T}~~~,~~F_{ab}\sigma^{ab}~~,~~\mathcal{R}_{ab}\sigma^{ab}$ $u_{a}\mathfrak{a}_{b}\mathcal{R}^{ab}~~,~~\theta \mathcal{R}~~,~~\theta u^{a}u^{b}\mathcal{R}_{ab}~~,~~\theta^{3}~~,~~\sigma^{ab}\sigma_{ab}\theta~~,~~\omega^{ab}\omega_{ba}\theta~~,~~\mathfrak{a}_{c}\mathfrak{a}^{c}\theta~~,~~\mathfrak{a}_{a}\mathfrak{a}_{b}\sigma^{ab}~~,~~\sigma_{ac}{\sigma^{c}}_{b}\sigma^{ba}$ $\omega_{ac}{\sigma^{c}}_{b}\omega^{ba}$ \\ \hline
     \end{tabular}
         \label{1order} 
\end{center}
\vskip 0.3cm
This completes the classification of the relevant independent structures appearing in the divergence of the entropy current.

\section{Another choice of surface for elastic corrections} \label{basis}
In Sec.~\ref{confined} we mentioned that particular choices of surfaces can eliminate certain terms in the bending moment. Our choice of surface consisted in not considering terms in the bending moment $\mathcal{D}^{abi}$ of the form $u^{a}u^{b}K^{i}$. Here we consider another choice of surface in which terms of the form $\gamma^{ab}K^{i}$ do not appear in the bending moment. Therefore we wish to consider a contribution of the form
\beq \label{last1}
\mathcal{D}^{ab}=\vartheta_2 u^{a}u^{b}~~,
\eeq
in the case of codimension-1 surfaces and
\beq \label{last2}
\mathcal{D}^{abi}=\lambda_4 u^{a}u^{b}K^{i}~~,
\eeq
in the case of codimension higher than one. Below, we consider such terms and compare the results with those arising from equilibrium partition functions. We then consider a specific combination of such terms that gives rise to the elasticity tensor. Finally, we write down the tensor $B^{ab}$ in the Landau gauge.

\subsection{Divergence of the entropy current and comparison with equilibrium partition functions}
Here we first consider the case of codimension-1 surfaces and then of higher codimension.
\subsubsection*{Codimension-1 surfaces}
For codimension-1 surfaces we consider the effect of adding a term of the form \eqref{last1}. Computing the divergence we find
\beq \label{div1_2}
\begin{split}
\nabla_{a}J^{a}_{s}|_{\vartheta_2\text{elastic}}=&~\left(-\frac{\kappa_1}{\mathcal{T}}+\pi_1-s\frac{\partial \pi_1}{\partial s}-\frac{s}{\mathcal{T}}\frac{\partial}{\partial s}\left(\frac{P}{\mathcal{T}s}\right)\vartheta_2+2\frac{s}{\mathcal{T}}\frac{\partial}{\partial s}\left(\frac{P}{\mathcal{T}s}\vartheta_2\right)-\frac{2}{\mathcal{T}}\frac{P}{\mathcal{T}s}\vartheta_2 \right)K\theta\\
&+\left(\pi_1-\frac{P}{\mathcal{T}s}\frac{\vartheta_2}{\mathcal{T}}\right)u^{a}\nabla_{a}K~~.
\end{split}
\eeq
Each of these terms is composed of independent fluid-elastic data and hence must vanish separately. Thus we find
\beq
\pi_1|_{\vartheta_2}=\frac{P}{\mathcal{T}s}\frac{\vartheta_{2}}{\mathcal{T}}~~,
\eeq
\beq
\kappa_1|_{\vartheta_2}=\mathcal{T}\pi_1-\mathcal{T}s\frac{\partial \pi_1}{\partial s}-s\frac{\partial}{\partial s}\left(\frac{P}{\mathcal{T}s}\right)\vartheta_2+2s\frac{\partial}{\partial s}\left(\frac{P}{\mathcal{T}s}\vartheta_2\right)-2\frac{P}{\mathcal{T}s}\vartheta_2~~.
\eeq
We now wish to compare this with the corresponding equilibrium partition function. The relevant action is of the form
\beq
I[X^{\mu}]=\int_{\mathcal{W}_{p+1}}\sqrt{-\gamma}\left(P+\tilde\vartheta_2 u^{a}u^{b}K_{ab}\right)~~.
\eeq
The stress-energy tensor and bending moment read \cite{Armas:2013hsa}
\beq
T^{ab}=T^{ab}_{(0)}-\frac{P}{\mathcal{T}s}K\left(\tilde\vartheta_2\gamma^{ab}-\left(\frac{\tilde\vartheta_2}{\textbf{k}^2}\right)'\textbf{k}^3 u^{a}u^{b}\right)~~,~~\mathcal{D}^{ab}=\tilde\vartheta_2u^{a}u^{b}~~,
\eeq
while the entropy current reads \cite{Armas:2014rva}
\beq
J^{a}_{s}=su^{a}-\frac{P}{\mathcal{T} s}\left(\textbf{k}^2\frac{\partial}{\partial\mathcal{T}}\left(\frac{\tilde\vartheta_2}{\textbf{k}^2}\right)'-2\frac{\tilde\vartheta_2}{\mathcal{T}}\right)Ku^{a}~~.
\eeq
By performing a frame transformation with parameters 
\beq
\delta\mathcal{T}=\frac{1}{\mathcal{T}}\frac{P}{\mathcal{T}s}\frac{\partial\mathcal{T}}{\partial s}\left(\tilde\vartheta_2+\mathcal{T}\frac{\partial\tilde\vartheta_2}{\partial\mathcal{T}}\right)K~~,~~\delta u^{a}=0~~,
\eeq
the stress-energy tensor and entropy current are brought to the form
\beq
\tilde{T}^{ab}=T^{ab}_{(0)}+\frac{P}{\mathcal{T}s}K\left(-\vartheta_2+\frac{s}{\mathcal{T}}\frac{\partial\mathcal{T}}{\partial s}-s\frac{\partial \tilde\vartheta_2}{\partial s}\right)P^{ab}~~,~~\tilde{J}^{a}_s=su^{a}+\frac{P}{\mathcal{T}s}\frac{\tilde\vartheta_2}{\mathcal{T}}Ku^{a}~~.
\eeq
These are in exact agreement with the results obtained above from the entropy current provided one identifies 
\beq
\vartheta_2=\tilde\vartheta_2~~.
\eeq

\subsubsection*{Codimension higher than one}
For codimension higher than one we consider the effect of a term of the form \eqref{last2}. To analyze this case we only need to consider the terms appearing in the stress-energy tensor \eqref{genst2} and entropy current \eqref{genE2} proportional to $\alpha_1$ and $\beta_1$. The divergence of this piece reads
\beq \label{div2_4}
\begin{split}
\nabla_{a}J^{a}_{s}|_{\lambda_4\text{elastic}}=&~\left(-\frac{\alpha_1}{\mathcal{T}}+\beta_1-s\frac{\partial \beta_1}{\partial s}-\frac{s}{\mathcal{T}}\frac{\partial}{\partial s}\left(\frac{P}{\mathcal{T}s}\right)\lambda_4+2\frac{s}{\mathcal{T}}\frac{\partial}{\partial s}\left(\frac{P}{\mathcal{T}s}\lambda_4\right)-\frac{2}{\mathcal{T}}\frac{P}{\mathcal{T}s}\lambda_4\right)\theta K^{i}K_{i} \\
&+ \left(-\frac{3}{\mathcal{T}}\frac{P}{\mathcal{T}s}\lambda_4+2\beta_1\right)u^{a}K^{\rho}\nabla_{a}K_{\rho}~~.
\end{split}
\eeq
Each of the above terms are proportional to linear independent fluid-elastic data and hence must be set to zero independently. Solving for the constraints we find
\beq
\beta_1|_{\lambda_4}=\frac{3}{2}\frac{P}{\mathcal{T}^2s}\lambda_4~~,
\eeq
and
\beq
\alpha_1|_{\lambda_4}=\mathcal{T}\beta_1-\mathcal{T}s\frac{\partial \beta_1}{\partial s}-s\frac{\partial}{\partial s}\left(\frac{P}{\mathcal{T}s}\right)\lambda_4+2s\frac{\partial}{\partial s}\left(\frac{P}{\mathcal{T}s}\lambda_4\right)-2\frac{P}{\mathcal{T}s}\lambda_4~~.
\eeq

We now wish to compare these results with those obtained in \cite{Armas:2013hsa, Armas:2014rva}. We consider an action of the form
\beq
I[X^{\mu}]=\int_{\mathcal{W}_{p+1}}\sqrt{-\gamma}\left(P+\tilde\lambda_4 u^{a}u^{b}u^{c}u^{d}{K_{ab}}^{i}K_{cdi}\right)~~.
\eeq
The stress-energy tensor and bending moment that follow from here are of the form \cite{Armas:2013hsa}
\beq
T^{ab}=T^{ab}_{(0)}+\left(\tilde\lambda_4 \gamma^{ab}-\left(\frac{\tilde\lambda_4}{\textbf{k}^4}\right)'\textbf{k}^5u^{a}u^{b}\right)u^{c}u^{d}u^{e}u^{f}{K_{cd}}^{i}K_{efi}~~,~~\mathcal{D}^{abi}=-2\frac{P}{\mathcal{T}s}\tilde\lambda_4 u^{a}u^{b}K^{i}~~,
\eeq
while the entropy current reads \cite{Armas:2014rva}
\beq
J^{a}_s=su^{a}+\left(\frac{P}{\mathcal{T}s}\right)^2\frac{\partial}{\partial\mathcal{T}}\left(\frac{\tilde\lambda_4}{\textbf{k}^4}\right)\textbf{k}^4K^{i}K_{i}u^{a}-4\frac{\tilde\lambda_4}{\mathcal{T}}u^{b}u^{c}u^{d}u^{e}{K_{bc}}^{i}{K_{dei}}u^{a}~~.
\eeq
In order to bring these structures to the Landau gauge we perform a frame transformation with parameters
\beq
\delta \mathcal{T}=-\frac{1}{\mathcal{T}}\left(\frac{P}{\mathcal{T}s}\right)^2\frac{\partial\mathcal{T}}{\partial s}\left(3\tilde\lambda_4+\mathcal{T}\frac{\partial \tilde\lambda_4}{\partial\mathcal{T}}\right)K^{i}K_{i}~~,~~\delta u^{a}=0~~,
\eeq
such that the stress-energy tensor becomes
\beq
\tilde{T}^{ab}=T^{ab}_{(0)}+\left(\frac{P}{\mathcal{T}s}\right)^2\left(\tilde\lambda_4-3\frac{s}{\mathcal{T}}\frac{\partial \mathcal{T}}{\partial s}\tilde\lambda_4-s\frac{\partial \tilde\lambda_4}{\partial s}\right)K^{i}K_{i}P^{ab}~~,
\eeq
while the entropy current reads \cite{Armas:2014rva}
\beq
\tilde J^{a}_s=su^{a}-3\left(\frac{P}{\mathcal{T}s}\right)^2\frac{\tilde\lambda_4}{\mathcal{T}}K^{i}K_{i}u^{a}~~.
\eeq
This is in complete agreement with the previous results obtained from requiring the divergence of the entropy current to vanish provided
\beq
\lambda_4=-2\frac{P}{\mathcal{T}s}\tilde\lambda_4~~.
\eeq

\subsection{The elasticity tensor}
The elasticity tensor arises naturally in a equilibrium partition function analysis as the contribution to the stress-energy tensor from terms that can be removed by a field redefinition as written in Sec.~\ref{confined}. Namely, from a contribution to the action of the form
\beq
I[X^{\mu}]=\int_{\mathcal{W}_{p+1}}\sqrt{-\gamma}\tilde k T^{ab}_{(0)}{K_{ab}}^{i}K_{i}~~,
\eeq
where $\tilde k$ is a function of $\mathcal{T}$. The contribution to the stress-energy tensor and bending moment are of the form
\beq
T^{ab}=\tilde{k}E^{abcd}{K_{cd}}^{i}K_{i}~~,~~\mathcal{D}^{abi}=\tilde{k}P\gamma^{ab}K^{i}+\tilde{k}\mathcal{T}su^{a}u^{b}K^{i}~~,
\eeq
where $E^{abcd}$ is the elasticity tensor introduced in Eq.~\eqref{elasticity}. We see that this gauge-variant contribution to the partition function is captured by our previous results provided we identify
\beq
\lambda_1=\tilde{k}P~~,~~\lambda_4=\tilde{k}\mathcal{T}s~~.
\eeq
Such terms can be removed from the stress-energy tensor and bending moment by performing a field redefinition of the form \eqref{varst2} with parameter
\beq
\tilde\varepsilon^{i}=\tilde{k}K^{i}~~,
\eeq
in agreement with \cite{Armas:2013hsa}.

\subsection{The tensor $B^{ab}$ in the Landau gauge}
As mentioned in Sec.~\ref{discussion}, when performing a measurement from gravity of the several transport coefficients what is measured directly is $B^{ab}$ as defined in Eq.~\eqref{bab}. Therefore, in the case of codimension-1 surfaces and performing the following frame transformation such that $B^{ab}$ is in the Landau gauge, 
\beq
\delta\mathcal{T}=\frac{2}{\mathcal{T}}\frac{P}{\mathcal{T}s}\frac{\partial \mathcal{T}}{\partial s}\vartheta_1 K~~,~~\delta u^{a}=\frac{2}{\mathcal{T}s}\vartheta_1{P^{a}}_{c}u_{d}K^{cd}~~,
\eeq
we find
\beq
B^{ab}|_{(1)\text{elastic}}=\left(\vartheta_1+\frac{s}{\mathcal{T}}\frac{\partial \mathcal{T}}{\partial s}\vartheta_1-s\frac{\partial\vartheta_1}{\partial s}\right)KP^{ab}~~,
\eeq
and the corresponding entropy current
\beq
J^{a}_{s}|_{(1)\text{elastic}}=\frac{\vartheta_1}{\mathcal{T}}Ku^{a}~~.
\eeq
In the case of codimension higher than one, the following frame transformation,
\beq
\delta\mathcal{T}=\frac{1}{\mathcal{T}}\frac{\partial\mathcal{T}}{\partial s}\left(\left(\frac{2}{\mathcal{T}}\frac{P}{\mathcal{T}s}\lambda_1-\left(\frac{P}{\mathcal{T}s}\right)^2\lambda_3\right) K^{i}K_{i}-\left(2\lambda_2-\lambda_3\right) u_{b}u_{d}K^{dci}{K^{b}}_{ci}\right)~~,
\eeq
\beq
\delta u^{a}=\frac{1}{\mathcal{T}s}{P^{a}}_{c}\left(\left(2\lambda_1 -\lambda_3\right)u_{d}K^{cdi}K_{i}+\left(2\lambda_2 -\frac{\lambda_3}{2}\right)u_{b}K^{cdi}{K^{b}}_{di}\right)~~,
\eeq
brings the tensor $B^{ab}$ to the Landau gauge and reads
\beq
\begin{split}
B^{ab}|_{(2)\text{elastic}}=&~\left(\left(\alpha_1+\frac{2}{\mathcal{T}}\frac{P}{\mathcal{T}s}\frac{\partial \mathcal{T}}{\partial s}\lambda_1\right) K^{i}K_{i}+\alpha_2 K^{abi}K_{abi}\right)P^{ab}\\
&+\left(\alpha_3-2\frac{s}{\mathcal{T}}\frac{\partial\mathcal{T}}{\partial s}\lambda_2+\frac{s}{\mathcal{T}}\frac{\partial\mathcal{T}}{\partial s}\lambda_3\right) u^{c}u^{d}{K_{c}}^{fi}{K_{dfi}}P^{ab}~~,
\end{split}
\eeq
while the entropy current in this gauge reads
\beq
\begin{split}
J^{a}_{s}|_{(2)\text{elastic}}=&~\left(\beta_1 K^{i}K_{i}+\beta_2 K^{cbi}K_{cbi}+\left(\beta_3+\frac{3}{2}\frac{\lambda_3}{\mathcal{T}}\right) u^{c}u^{d}{K_{c}}^{fi}K_{dfi}\right)u^{a} \\
&+\frac{\lambda_3}{2\mathcal{T}} u_{c}K^{abi}{K^{c}}_{bi}~~.
\end{split}
\eeq

\addcontentsline{toc}{section}{References}
\footnotesize
\providecommand{\href}[2]{#2}\begingroup\raggedright\endgroup

\end{document}